\documentclass{aa}
\usepackage{graphicx}
\usepackage{txfonts}
\usepackage{threeparttable}
\usepackage{hyperref}

\def\CNC{55 CNC e\xspace}
\def\OT{\mathrm{O_2}\xspace}
\def\O{\mathrm{O}\xspace}
\def\H{\mathrm{H}\xspace}
\def\K{\mathrm{K}\xspace}
\def\Na{\mathrm{Na}\xspace}
\def\Mg{\mathrm{Mg}\xspace}
\def\Fe{\mathrm{Fe}\xspace}
\def\Si{\mathrm{Si}\xspace}
\def\HT{\mathrm{H_2}\xspace}
\def\HTO{\mathrm{H_2O}\xspace}
\def\SiOT{\mathrm{SiO_2}\xspace}
\def\SiO{\mathrm{SiO}\xspace}
\def\SiHF{\mathrm{SiH_4}\xspace}
\def\SiHT{\mathrm{SiH_3}\xspace}

\begin{document}

   \title{The effect of a small amount of hydrogen in the atmosphere of ultrahot magma-ocean planets: atmospheric composition and escape  }

   \titlerunning{Hydrogen above a magma ocean}
  \authorrunning{S. Charnoz et al.}
   


   \author{Sébastien Charnoz\inst{1}, Aurélien Falco \inst{1,2}, Pascal Tremblin \inst{3}, Paolo Sossi \inst{4}, Razvan Caracas \inst{1,5}, Pierre-Olivier Lagage \inst{2}
          }

   \institute{(1) Universit\'e de Paris Cité, Institut de Physique du Globe de Paris, CNRS
F-75005 Paris, 1 rue Jussieu, 75005 Paris, France\\
              \email{charnoz@ipgp.fr}\\
    (2) Laboratoire AIM, CEA, CNRS, Univ. Paris-Sud, UVSQ, Université Paris-Saclay, F-91191 Gif-sur-Yvette, France  \\
    (3) Université Paris-Saclay, UVSQ, CNRS, CEA, Maison de la Simulation, 91191, Gif-sur-Yvette, France \\
    (4) Institute of Geochemistry and Petrology, ETH Zürich, CH-8092, Zürich, Switzerland \\
    (5) CEED, PHAB, University of Oslo, Oslo, Norway
             }

   \date{Received September 15, 1996; accepted March 16, 1997}


  \abstract
  {Ultrahot (> 1500K) rocky exoplanets may be covered by a magma ocean, resulting in the formation of a vapor rich in rocky components (e.g., Mg, Si, Fe) with a low total pressure and high molecular mass. However, exoplanets may have also captured a significant amount of hydrogen from the nebular gas during their formation. Ultrahot rocky exoplanets around the Fulton gap ($\sim 1.8 R_\oplus$) are sufficiently large to have retained some fraction of their primordial hydrogen atmosphere.}
  { Here we investigate how small amounts of hydrogen (much smaller than the mass of the planet) above a magma ocean may modify the atmospheric chemistry and its tendency to thermally escape.}
   {We use a chemical model of a magma ocean coupled to a gas equilibrium code (that includes hydrogen)
   to compute the atmospheric composition at thermodynamical equilibrium for various H contents and temperatures. An energy-limited model is used to compute atmospheric escape and is scaled to consider H-rich and H-poor atmospheres. }
  {The composition of the vapor above a magma ocean is drastically modified by hydrogen, even for very modest amounts of H ($\ll 10^{-6}$ planetary mass). Hydrogen consumes much of the O$_2$(g), which, in turn, promotes the evaporation of metals and metal oxides (SiO, Mg, Na, K, Fe) from the magma ocean. Vast amounts of $\HTO$ are produced by the same process. At high hydrogen pressures, new hydrogenated species such as $\SiHF$ form in the atmosphere. In all cases, H, H$_2$,  and $\HTO$ are the dominant nonmetal-bearing volatile species. Sodium is the dominant atmospheric metal-bearing species at T$<$ 2000K and low H content, whereas Fe is dominant at high H content and low temperature, while SiO predominates at T>3000 K.  We find that 
 the atmospheric Mg/Fe, Mg/Si, and Na/Si ratios deviate from those in the underlying planet and from the stellar composition. As such, their determination may constrain the planet's mantle composition and H content. As the presence of hydrogen promotes the evaporation of silicate mantles, it is conceivable that some high-density, irradiated exoplanets may have started life as hydrogen-bearing planets and that part of their silicate mantle evaporated (up to a few $10 \%$ of Si, O, and Fe) and was subsequently lost owing to the reducing role of H.}
  {Even very small amounts of H can alter the atmospheric composition and promote the evaporation to space of heavy species derived from the molten silicate mantle of rocky planets. Through transit spectroscopy, the measurement of certain elemental ratios, along with the detection of atmospheric water or hydrogen, may help to determine the nature of a surface magma ocean. }

   \keywords{exoplanets, magma oceans, hydrogen}

   \maketitle
%
\section{Introduction}

There are more than 350 ultrahot exoplanets (with equilibrium temperature of > 1500~K) discovered as of July 2022. About half of them have radii of < 5 Earth radii and have masses that are consistent with either pure rocky planets or sub-Neptunes. At such temperatures and assuming a silicate-based mantle, their surfaces should be partially or entirely covered by a magma ocean . Such planets are of particular interest for future spectroscopic observations because convection in the mantle should be sufficiently rapid \citep{solomatov2000} such that their atmospheres are readily replenished by chemical exchange with the underlying molten surface. As such, any spectroscopic constraints on the nature of their atmospheres may provide insights into the composition of their molten mantles. However, the search for atmospheres on ultrahot rocky worlds has so far brought mostly negative results. Spectroscopic observation of LHS 3844b (with an equilibrium temperature $\sim$ 1000K) yielded results that are compatible with the absence of an atmosphere \citep{Kreidberg_2019}. The detection of an atmosphere around \CNC is debated. Although a hotspot offset has been reported in the planet's phase curve  \citep{Demory_2016} ---which can be interpreted as the consequence of a thick and opaque atmosphere---, detailed spectroscopic observations of \CNC have not  unambiguously detected any gaseous species \citep{Bourrier_2018_55CNCE}. Some hints of the possible presence of HCN were reported in \cite{Tsiaras_2016}, but this would seem unlikely on chemical equilibrium grounds given that hydrogen itself has never been detected (see e.g., \cite{Ehrenreich_2012}). 
   
\begin{figure}[ht]
   \includegraphics[scale=0.25]{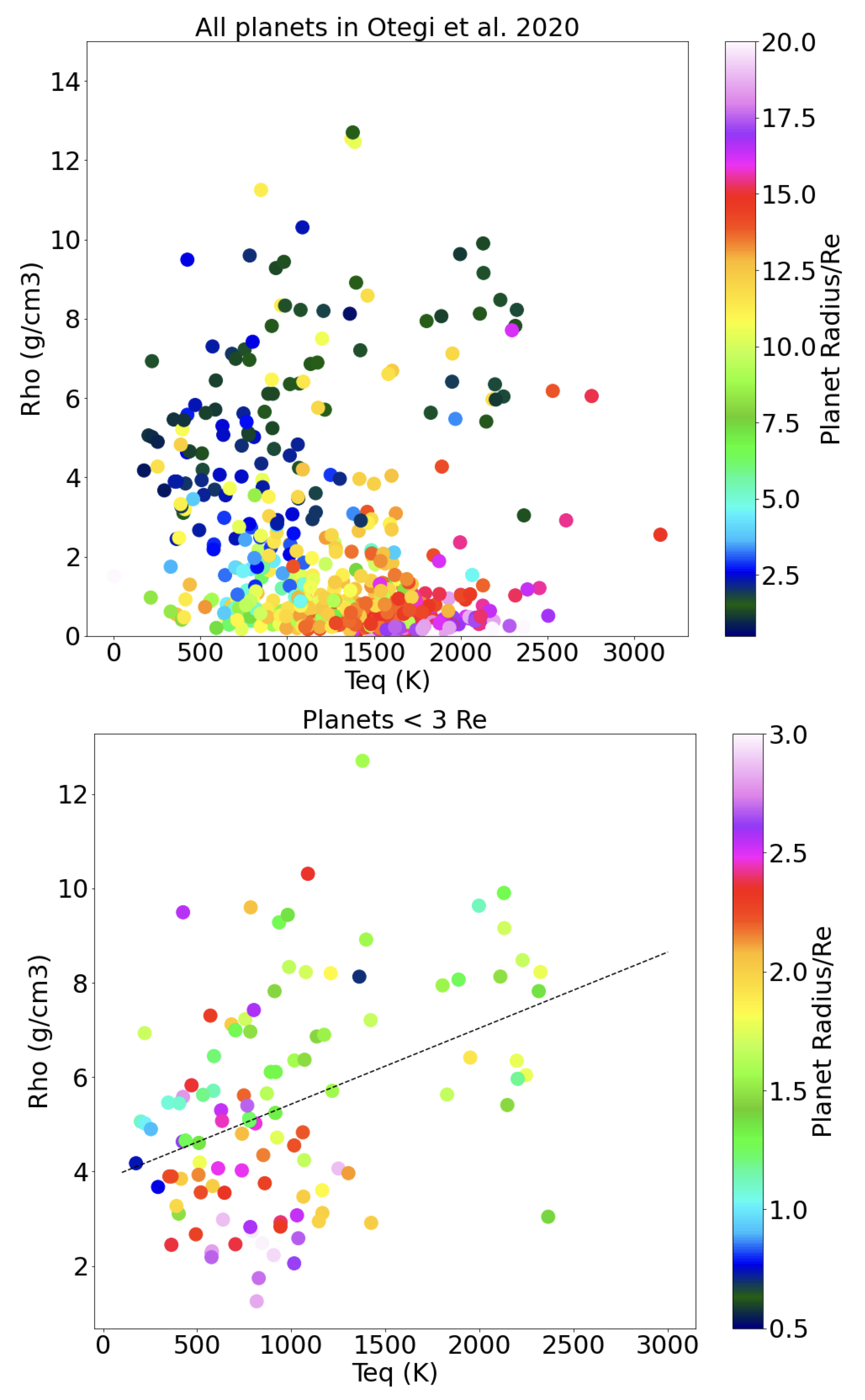}
   \caption{ Density of known exoplanets versus their equilibrium temperatures. These data come from the online DACE database maintained by Observatory of Geneva\textsuperscript{1} and are taken from the catalog of \cite{Otegi_2020}, which is an exoplanet catalog based on reliable, robust, and ---to the greatest possible extent--- accurate mass and radius measurements of transiting planets up to 120 earth masses  \citep{Otegi_2020}.  (Top) Density vs. equilibrium temperature for all planets. (Bottom) Same as above but limited to planets with radii of smaller than 3 Earth radii. The dotted line shows a simple least-square fit to the data. }
   \scriptsize\textsuperscript{1}~\url{https://dace.unige.ch/}
    \label{Fig_density_VS_Teq}%
\end{figure}

Recently, \cite{Zieba_2022} reported the possible indirect detection of an atmosphere around the ultrahot planet ($T_{eq}\sim~2000$~K) K2-141b by comparing Spitzer to synthetic spectra. However, the resolution of Spitzer spectra is too low to permit any clear identification of the molecular species in the partial pressures expected above such a rocky planet.
Therefore, the detection of any sort of atmosphere around a lava planet remains elusive, despite growing international research efforts towards this goal. For the sake of completeness, we mention that a low-mean-density atmosphere containing HCN and CH$_4$ has been detected around the rocky world GJ 1132, but it is a "warm" planet (T$\sim$ 530~K), and may not have a surface magma ocean \citep{Swain_2021}.
Interestingly, ultrahot rocky planets tend to have high average densities. For example, 55 Cnc e has an estimated density of $\sim$ 5.9~g/cm$^3$ \citep{Crida_2018_55CNCe, Mercier_2022_55CNCe}.  When comparing the density of exoplanets with their equilibrium temperatures (Fig.  \ref{Fig_density_VS_Teq} bottom), strongly irradiated rocky planets with equilibrium temperatures of T$_{eq}$ > 1700~K  have, on average, a much higher density than those planets with T$_{eq}$ below 1700~K. 
   
\begin{figure}
   \centering
   \includegraphics[scale=0.1]{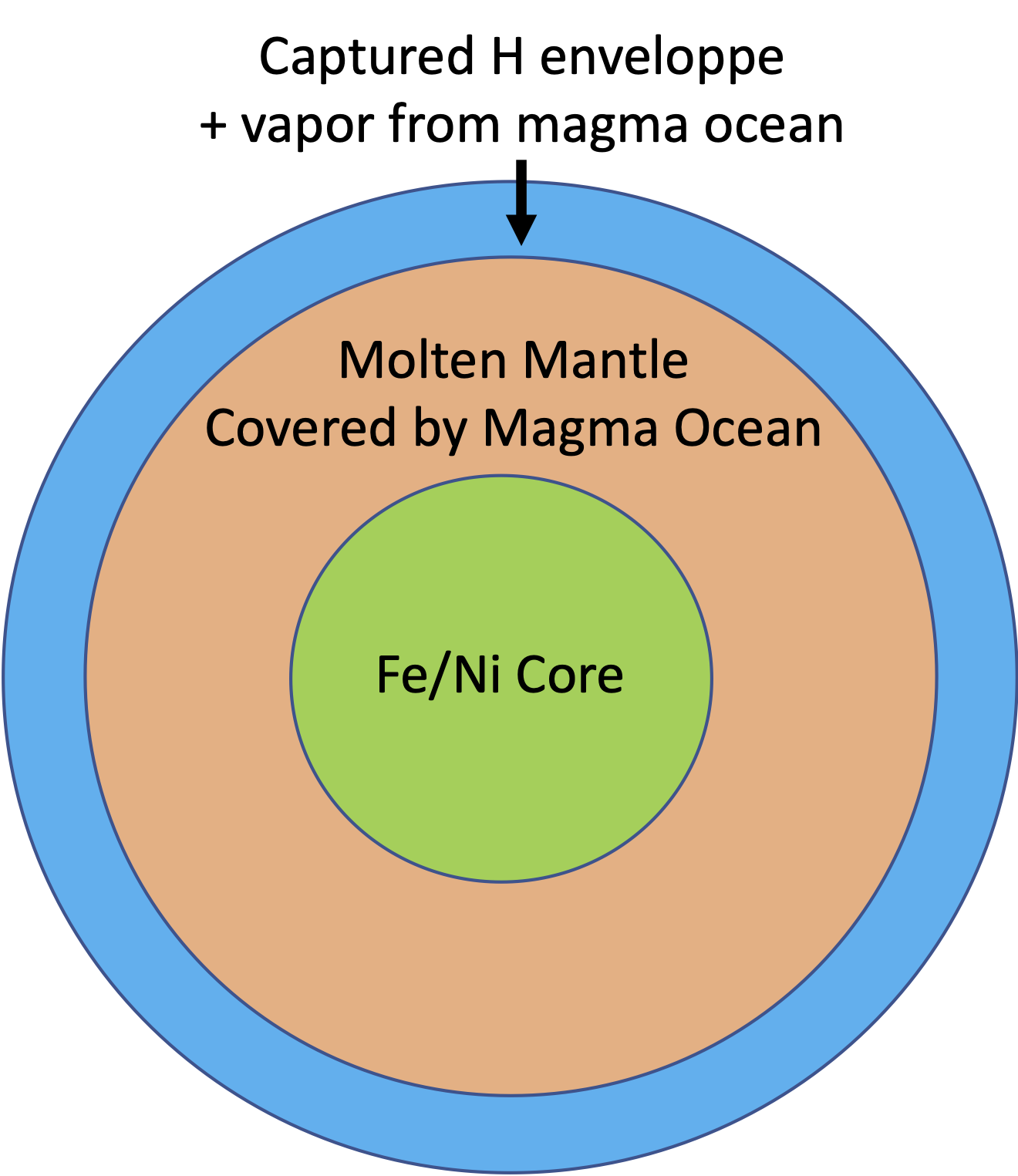}
   \caption{ Schematic illustration of the internal structure of a rocky planet as assumed in this paper, with an atmosphere containing a mixture of captured hydrogen and gaseous species derived from a molten mantle that forms a global magma ocean below the planet atmosphere. In the centre of the planet, an iron/nickel core is present. The composition of the magma ocean is assumed to be fixed, while the composition of the atmosphere is at thermodynamic equilibrium with that of the magma ocean.}
    \label{Fig_planet_internal_structure}%
\end{figure}

   More specifically, the population of terrestrial planets with densities < 5~g/cm$^3$ almost completely disappears at T$_{eq}$ larger than 1700~K (Fig.  \ref{Fig_density_VS_Teq}), and all observed ultrahot rocky planets smaller than 3 Earth radii have densities ranging from 5 to 11 g/cm$^3$(with only one remarkable exception of a planet with T$_{eq}\sim~2400$~K and $\rho \sim 3$~g/cm$^3$).  \cite{Otegi_2020} demonstrate that volatile-rich and rocky exoplanet populations may be separated on the basis of their density, with a transition around 3~g/cm$^3$.  Ultrahot mini-Neptunes and terrestrial planets may have a thin- to moderately massive atmosphere  (less than 50\% of the planet's mass) above a molten  mantle and an iron-rich core, as sketched out in Figure \ref{Fig_planet_internal_structure}.  
The broad positive trend of density versus equilibrium temperature might reflect a decreasing mantle/core mass ratio with increasing surface temperature. This trend could either result from efficient iron condensation (relative to silicates) in the hottest region of the protoplanetary disk during the formation of first solids (see e.g., \cite{Pignatale_2017,Johansen_2022_Iron_condensation}) or arise from collisional \citep{oneillpalme2008, denman2020} or evaporative \citep{young_etal2019, charnoz_etal2021, jaggi_etal2021} mass loss. However, most terrestrial planets are thought to have moved inward during their formation due to efficient  migration, and so highly irradiated magma ocean exoplanets could have formed further away at much lower temperatures and with much higher hydrogen budgets than they have today. \\
   
In the present paper, we wish to investigate the possibility that the mantles of rocky planets in a magma ocean state evaporate at high temperatures in the presence of H$_2$ gas. Depending on the efficiency of this process, it might lead to a decrease in the mantle mass with increasing temperature, and thus an increase in planet density, given that the iron-rich core remains isolated from the evaporation occurring at the surface. Although the general treatment is not new, and evaporation from a magma ocean has been studied under varying circumstances \citep[see e.g.,][]{Schaefer_2010_atmo_formation,Lupu_2014_atmo_after_GI,Schaefer_2012,Ito_2015,Ito_Ikoma_2021, charnoz_etal2021, jaggi_etal2021}. It was found that the vapor above a magma ocean with a composition like that of Earth's mantle (also called bulk silicate Earth, or BSE) has, on an anhydrous basis (i.e., neglecting H, C, Cl, S, N, and other volatiles) very low pressure ($<< 1$ bar in general) and is dominated by Na and K for T~<~2500~K, and by SiO for T~>~3000~K \citep{Schaefer_2010_atmo_formation,Schaefer_2012,Ito_2015,Ito_Ikoma_2021}. 
The escape of a mineral atmosphere was studied in detail by \cite{Ito_Ikoma_2021}, and was found to be very inefficient.  A mineral-based atmosphere released by a magma ocean  is rich in sodium, potassium, oxygen, iron, magnesium, and silicon \citep{Schaefer_2010_atmo_formation}.
\cite{Schaefer_Fegley_2004} show that the dominant metallic gas above a silicate melt should be SiO for  $T>2700$~K and Na for lower temperature. \cite{Ito_Ikoma_2021} show that this mineral vapor dissociates to simple atomic and ionic compounds in the upper atmosphere. Because Na and K are efficient atmospheric coolants \citep{Ito_Ikoma_2021}, most of the energy transferred to the planet through irradiation by 
the star is re-emitted to space, making thermal escape inefficient with resulting heating efficiency in the range of $\sim 5x10^{-4}$ to  $\sim 5x10^{-3}$, compared to 0.14 to 0.4 for hydrogen-dominated atmosphere \citep{Valencia_2015}. \cite{Ito_Ikoma_2021} therefore concluded that atmospheric escape would have only a marginal influence on the composition of ultrahot rocky planets covered by a magma ocean. Nonetheless, \citet{Ito_Ikoma_2021}  note that, at 3000~K, most of the Na and K content of the planet's mantle should escape during the lifetime of the star. However, as Na and K may only represent a very small proportion of the mantles of such planets (together they comprise < 1 mol $\%$ in the BSE for example), even if the entire planetary budget of Na and K were to escape, their loss would be unlikely to significantly affect either the mass or the radius of the rocky planet. For the atmospheric escape of the magma to have a significant impact on the mass or radius of the planet, major components, such as O, Mg, or Si, must be released in the vapor in very high quantities and then lost to space.

Here, we revisit the idea that atmospheric escape in the presence of hydrogen can engender a significant loss of the  mantle of a rocky planet. The specificity of our approach lies in our consideration of the presence of a hydrogen envelope above the magma ocean and in computation of the consequences this has on the atmosphere's speciation (which is modeled as a mixture between the H captured from the protoplanetary disk and species evaporated from the magma ocean) and atmospheric loss.  To date, several studies have investigated the dissolution of H$_2$ or H$_2$O in a silicate magma ocean, and the effect on the planet's mass and radius  atmospheric loss (see e.g., \cite{Kite_2020, Kite_Schaefer_2021}), but neither the change in atmospheric speciation nor the identities and partial pressures of the major rock-forming elements have been investigated.

To our knowledge, the only other attempt to characterize the effect of H on the composition of a hot rocky planet atmosphere was by \cite{Fegley_2016_steam_atmo}, who computed (based on the thermodynamic properties of hydroxide- and halide gas species) the compositions of metal-bearing atmospheres above a BSE magma for various partial pressures of H$_2$O, but did not consider the case of a captured  H$_2$ envelope above a magma ocean. Although the effect of H$_2$ has been examined in its influence on the escape of K from rocky planets \citep{erkaev_etal2022}, these authors neglect H solubility in the magma ocean and do not provide a self-consistent model of atmospheric speciation in equilibrium with the magma. 
\cite{Lichtenberg_2021, Dorn_2021} considered the effect of dissolved $\HT$ and $\HTO$ on the structure of an exoplanet, but did not compute the effect of H on the equilibrium chemistry of the atmosphere. On the other hand, \cite{sossi2020redox}, \cite{bower2022retention}, \cite{gaillard2022redox}, \cite{sossi2023solubility} determine the equilibrium speciation of an atmosphere in equilibrium with magma oceans for a range of H/C, \textit{f}O$_2$, and H masses, and argue that oxidized conditions promote the dissolution of H$_2$O in the magma ocean. However, these studies did not consider the influence of H$_2$O or H$_2$ on the speciation of rock-forming elements in the gas phase. 


Because H$_2$ is the dominant constituent of protoplanetary disks, it may be captured during the planet formation process, even around growing rocky planets \citep[e.g.,][]{OlsonSharp2019}. The ubiquity of H may therefore play a major role in influencing the chemistry of atmospheres degassing from rocky planets. We concentrate our efforts on examining the effect of H on the gaseous species present above silicate melts in the system Na-K-Mg-Al-Fe-Si-O-H. We find that the evaporation of metal-bearing species from a BSE-like magma ocean is strongly enhanced by the presence of hydrogen. It has been shown that a H$_2$-rich atmosphere can be lost rapidly for ultrashort period planets, in less than 10 Myr (see e.g., \cite{Lopez_2017,erkaev_etal2022}). Therefore, the effect of a captured H$_2$ envelope on atmospheric chemistry and its subsequent loss merits further investigation. In particular, we find that even minor amounts of H (with respect to the planet's mass) may have a strong influence on the atmosphere composition.


In the first part of the present paper, we aim to quantify how the presence of H modifies the vapor composition compared to a H-free case using a thermodynamic model of vapor--magma ocean equilibrium. To illustrate the physical processes at play, in section \ref{sec:toy_model} we present a simplified model based on two reactions that show how H modifies liquid--vapor equilibria in promoting the vaporization of metal-bearing gases due to the decrease in oxygen fugacity (=~O$_2$ partial pressure) it engenders, as well as the stabilization of H-bearing metal molecules in the gas phase. This leads to extensive depletion of Si and other metals from the silicate liquid. In section \ref{sec_method_full_model}, we consider a more realistic case of an infinite magma ocean with BSE composition in contact with a hydrogen envelope. We compute the vapor composition as a function of temperature and the amount of captured hydrogen.
At the end of this section, we also show that measurements of Na/Si, Mg/Si, and Mg/Fe and the detection of H$_2$O in the atmosphere of ultrahot rocky planets may be used as indicators of the presence and nature of a magma ocean.
  
 The second part of the paper explores 
 how the presence of H promotes atmospheric escape of heavy metallic species by increasing their concentration in the atmosphere (SiO, O, O$_2$, Fe, Mg, etc.). Should this process prevail over long timescales, this may lead to considerable evaporation of mantle components (relative to the Fe-Ni-rich core), and thus give rise to high-density rocky planets. We use a simple energy-limited formalism modified to take into account the inefficient loss of mineral species as reported in \cite{Ito_2015} and in \cite{Ito_Ikoma_2021} to quantify the extent of atmospheric loss. 
 

\section{Toy model: A $\SiOT$ magma ocean degassing in a $\HT$ atmosphere}\label{sec:toy_model}
We start with a simple chemical model to illustrate the effect of the presence of a primordial hydrogen envelope at the surface of a degassing magma ocean. For simplicity, we assume that the (infinite) magma ocean is made of $\SiOT$ only at temperature T and is  surrounded by a $\HT$ atmosphere. To understand the basic ingredients, we reduce the physics to two equations (1) the vaporization of $\SiO_{2(\ell)}$ (where the index $(\ell)$ stands for liquid and $(g)$ for gas) in $\SiO_{(g)}$ and $\OT$, and (2) the conversion of $\HT$ and $\OT$ into $\HTO$. These reactions are

      \begin{equation}
          \SiO_{2(\ell)} \Leftrightarrow \SiO_{(g)}+1/2 \OT,
          \label{reac_vapo_SIO2}
      \end{equation}

       \begin{equation}
          \HTO \Leftrightarrow \HT+1/2 \OT.
         \label{reac_H2O}
       \end{equation}


At equilibrium between the liquid and the gas, the partial pressure of $\SiO_{(g)}$ is given by the law of mass action at temperature $T$:

    \begin{equation}
      P_{\SiO_{(g)}}P_{\OT}^{1/2} = a[\SiO_{2(\ell)}] K_1(T),
     \label{equil_SIO2}
    \end{equation}

where $P_{\SiO_{(g)}}$ is the partial pressure of $\SiO_{(g)}$, $a[\SiO_{2(\ell)}]$ is the activity of $\SiOT$ in the melt (here we consider an ideal and pure liquid so that $a[\SiO_{2(\ell)}]=1$ here) and $P_{\OT}$ is the partial pressure of $\OT$, that is the $f\OT$ (we consider an ideal gas, such that the fugacity, $f$ , is equal to the partial pressure, $p$). Here, $K_1(T)$ is the equilibrium constant of this reaction at temperature T. The equilibrium constant is then related to the Gibbs-free-energy change of the evaporation reaction,
$K_1(T)=e^{-\Delta G1/RT}$ where $\Delta G1=\Delta G^0(\SiO_{(g)})+1/2 \Delta G^0(\OT)-\Delta G^0(\SiO_{2(\ell)})$ and $\Delta G^0$ are the Gibbs free energies of formation of the different species from the elements. Similarly, for the $\HTO$ formation equation (Eq.~\ref{reac_H2O}):

    \begin{equation}
      P_{\OT}^{1/2}P_{\HT}= K_2(T) P_{\HTO}.
     \label{equil_H2O}
   \end{equation}

 To close the system, we introduce the following limiting conditions. The first is that all oxygen atoms that go in the atmosphere come from the dissociation of $\SiOT$, such that the atomic O:Si ratio in the atmosphere is always equal to 2, which leads to the relation:
 \begin{equation}
      P_{\SiO_{(g)}}+P_{\HTO}+2P_{\OT}=2P_{\SiO_{(g)}}.
     \label{equil_sum_O}
  \end{equation}
 The above equation is established by considering that the total number of oxygen moles in the system is $N_O=N_{\SiO}+2N_{\OT}+N_{\HTO}$, and that $N_\O=2N_{\SiO}$, if we assume that all oxygen comes from the dissociation of $\SiO_{2(\ell)}$ . We therefore get the equality :  $2N_{\SiO}=N_{\SiO}+2N_{\OT}+N_{\HTO}$. As partial pressures are proportional to the number of moles, one obtains $2P_{\SiO}=P_{\SiO}+2P_{\OT}+P_{\HTO} $. 
 In addition, we assume that the total H content of the atmosphere is fixed and is equal to $P^0_H$ , which is defined as:

 \begin{equation}
      P^0_H=2P_{\HTO}+2P_{\HT}
     \label{equil_sum_H}
 ,\end{equation}
 and is used as a proxy for the total quantity of H in the atmosphere in the remainder of the paper.
 Equations \ref{equil_SIO2} to \ref{equil_sum_H} form a close set of equations with unknowns ($P_{\OT}, P_{\HTO}, P_{\SiO_{(g)}}, P_{\HT}$) and free parameters ($T, P_\H^0$). We solve this simple system numerically. \\

\begin{figure}
   \includegraphics[width=.5\textwidth]{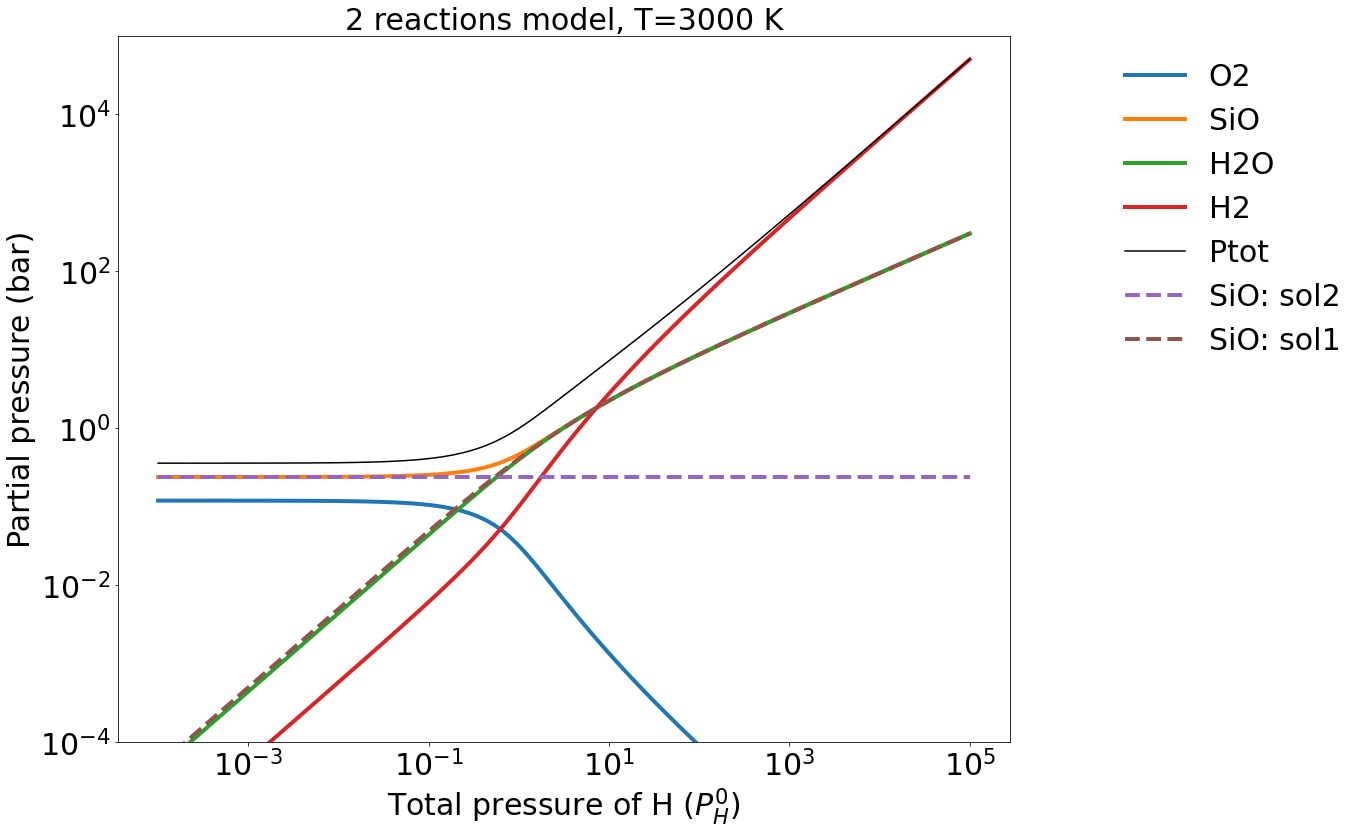}
   \caption{Pressures of the different elements for the simple two-reaction model with only $\SiO_{(g)}$, $\HT$, $\HTO,$ and $\OT$. Here, the temperature is fixed to 3000~K.  SiO solution 1 and SiO solution 2 refer to asymptotic solutions with low and high hydrogen content, respectively.
   }
    \label{Fig_simple_model_Tfixed}%
\end{figure}

\begin{figure}
   \includegraphics[width=.5\textwidth]{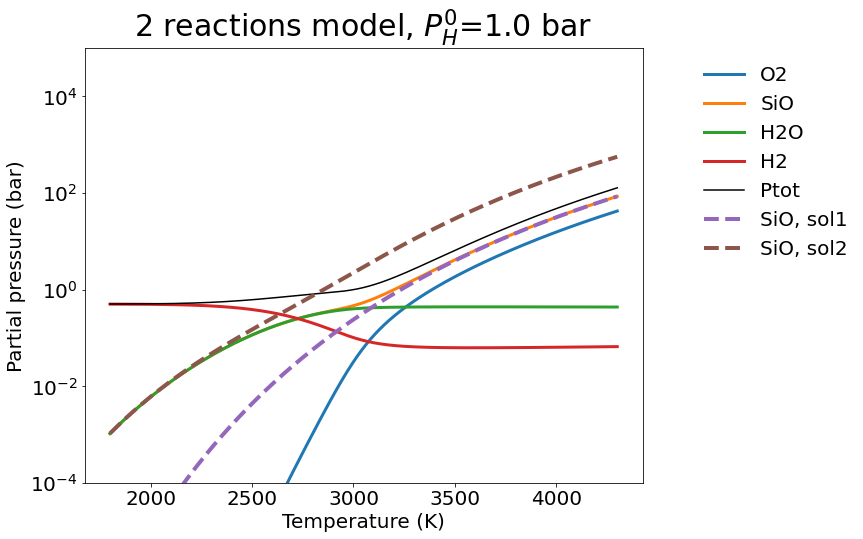}
   
   \caption{Pressures of the different elements for the simple two-reaction model with only $\SiO_{(g)}$, $\HT$, $\HTO,$ and $\OT$. Here, the hydrogen content is fixed ($P_\H^0=1$bar). SiO solution 1 and SiO  solution 2 refer to asymptotic solutions with low and high hydrogen content, respectively.
   }
    \label{Fig_simple_model_H}%
\end{figure}

 The partial pressures of the various gas species as a function of the initial H content and temperature are displayed in Figure \ref{Fig_simple_model_Tfixed}. For low values of $P_\H^0$ (< 1 bar), the atmosphere is dominated by $\OT$ and SiO$_{(g)}$ and corresponds to a pure rocky atmosphere in a hydrogen-free environment. As $P_\H^0$ increases, $\HTO$ begins to form, and $P_{\OT}$ drops sharply in response. As the system becomes increasingly oxygen-poor, the quantity of silicon in the atmosphere increases proportionally to $P_{O_{2}}^{-(1/2)}$, as given by Eq. \ref{equil_SIO2}. In Figure \ref{Fig_simple_model_H}, 
 the hydrogen content of the atmosphere $P_\H^0$ is set, while the temperature is varied. At low temperatures, H$_2$ predominates. However, as temperature increases, the equilibrium (2) shifts to the left, and H$_2$O becomes the prevailing H-bearing gas species above $\sim$ 2700 K. Above $\sim$ 3000 K, the H$_2$/H$_2$O ratio remains constant and SiO(g) and O$_2$(g) increase in tandem according to the $P_{O_{2}}^{-(1/2)}$ dependence.


We now consider two asymptotic cases. The partial pressure of $\SiO_{(g)}$ in the H-poor and H-dominated regimes can be easily computed as follows.
If the  pressure of hydrogen is extremely low, then $\mathrm{H_2O}$ and $\mathrm{H_2}$ are mostly absent and the system reduces to the single reaction written in Eq. \ref{reac_vapo_SIO2}.
 

Therefore, the relation between the pressures of $\SiO_{(g)}$ and $\OT$ is given by Eq.~\ref{equil_SIO2}. We set $a[\SiO_{2(\ell)}]=1$ (pure and ideal liquid). The dissociation of $\SiO_{2(\ell)}$ implies that there are always two atoms of O for 1 atom of Si in the gas. Counting the number of O and Si atoms in each species leads to the relation:
 \begin{equation}
    2P_{\OT_{(g)}}+P_{\SiO_{(g)}}=2 P_{\SiO_{(g)}}
    \label{eq_mass_bal_O_asympt_case_1}
.\end{equation}

Combining Equations \ref{equil_SIO2} and \ref{eq_mass_bal_O_asympt_case_1} and solving for $P_{\SiO_{(g)}}$ leads to the simple solution:

\begin{equation}
    P_{\SiO_{(g)}}(T)=2^{1/3}K_1(T)^{2/3}.
\end{equation}
This solution is displayed in Figure \ref{Fig_simple_model_Tfixed} with the brown dashed line. In the opposite case, in which hydrogen becomes a major component, $\HT$ and $\HTO$ control the system evolution. The system is therefore controlled by the reaction
\begin{equation}
          \SiO_{2(\ell)}+\HT \Leftrightarrow \SiO_{(g)}+\HTO.
          \label{reac_vapo_SIO2-H2O}
\end{equation}

The law of mass action gives the relationship between the partial pressures:

\begin{equation}
      P_{\SiO_{(g)}}P_{\HTO}= K_3(T) P_{\HT}a[\SiO_{2(\ell)}],
     \label{equil_SIO2-H2O}
\end{equation}

where
\begin{equation}
K3(T)=e^{-\frac{G^0(\SiO_{(g)})+G^0(\HTO)-G^0(\HT)-G^0(\SiO_{2(\ell)})}{RT}}= K_1(T)/K_2(T). 
\end{equation}

As all oxygen in the vapor comes from  the evaporation of $SiO2$, the conservation of the oxygen atoms gives the relation $P_{\HTO}+P_{SiO}=2P_{SiO}$, which reduces to $P_{\HTO}=P_{SiO}$. The conservation of the hydrogen atoms gives $2P_{\HTO}+2P_{\HT}=P^0_H$. We set $a[\SiO_{2(\ell)}]=1$ as usual. Solving for $P_{\SiO_{(g)}}$ leads to a second-order polynomial whose only positive solution is

\begin{equation}
      P_{\SiO_{(g)}}(P_\H^0, T)=\frac{-K_3+\left(K_3(T)^2+2K_3(T)P_\H^0 \right)^{1/2}}{2}
     \label{eq_PSIO_asymptot_solution_high_H}
.\end{equation}

This asymptotic solution is displayed as a purple dashed line in Figure \ref{Fig_simple_model_Tfixed}, and closely matches the SiO behavior for large hydrogen content. This also explains why SiO and $\HTO$ have the same values at high H pressure. We see that at high hydrogen pressure, $\HTO$ plays both the role of O and H reservoir, while $\OT$ is almost completely absent. This clearly demonstrates that, as the pressure of H increases, the content of Si in the atmosphere sharply increases also.

We conclude from this section that H plays a major in the composition of the vapor: because of the formation of $\HTO$, the O2 partial pressure drops and, in turn,   
we observe an enhanced evaporation of SiO that scales with $P_\H^{0^{1/2}}$. We now turn to the full model where this process acts on all metals contained in the magma ocean.

\section{Full model method calculation}
\label{sec_method_full_model}
Here, we describe a model of an infinite magma ocean (infinite = the liquid magma composition is fixed) in equilibrium with a H-rich primordial atmosphere. 
Our code is not unique; the MAGMA \cite{Fegley_Cameron_1987_MAGMA} and VapoRock \citep{Wolf_2022} codes perform similar calculations. In particular, the VapoRock code takes into account nonideal interactions via a Regular solution model. For the vapor molecular composition, our approach uses a Gibbs free-energy minimization code (we use the iconic CEA/NASA code provided by \cite{Gordon_McBride_1996_CEA_code}). We modify the CEA code to couple it to a vapor-liquid phase equilibrium model corresponding to liquid with a fixed composition. The same approach may be used with any atmospheric code \citep[e.g.,][]{Woitke_2018_GGCHEM}.  We assume that the magma ocean is made of a nonideal mixture of liquid oxides, namely $\SiO_{2(\ell)}, \Mg\O_{(\ell)},\Fe\O_{(\ell)}, \Na_2\O_{(\ell)}$, and $\K_2\O_{(\ell)}$, with fixed molar abundances $X^l_i$ (where $i$ stands for any of the liquid oxides). Each liquid oxide may vaporize into gaseous species through the limited list of reactions compiled in Table \ref{table:Table_reaction_liquid_gas}. The specificity of our approach compared to previous works is that we assume there is a pre-existing H envelope whose total pressure of monoatomic H is $P_\H^0$. For now, we assume that all H stays in the vapor phase, whereas it is known that $\mathrm{H_2}$ and $\mathrm{H_2O}$  may dissolve in the magma ocean \citep{Hirschmann_2012, sossi2023solubility}. This simplification is not critical here as $P_\H^0$ may be interpreted as the quantity of H contained in the atmosphere only. 

\subsection{The H monoatomic pressure: $P_\H^0$ }
The H envelope reacts with gaseous species derived from the magma ocean to modify the atmospheric composition. Therefore, knowledge of the total amount of H in the atmosphere is useful. In order to quantify the amount of H independently of the atmosphere composition and planet mass, we introduce the "hydrogen monoatomic pressure", $P_\H^0$, which corresponds to the partial pressure of H if all the H of the atmosphere is in pure monoatomic form, and is computed as follows:

\begin{equation}
    P_\H^0= \sum_{i}^{} \nu_H^iP_i,
     \label{PH0_def}
\end{equation}

where i is any molecule in the atmosphere, $P_i$ is the partial pressure of molecule $i$, and $\nu_H^i$ is the stochiometric coefficient of H in molecule $i$. For example, in an atmosphere in which all H is in $\HT$,  $P_\H^0=2\times P_{\HT}$ where $P_{\HT}$ is the partial pressure of $\HT$.

\subsection{Constitutive equations}
For a given value of $P_\H^0$ and magma temperature, the parameters we wish to determine are the molar fractions of each gaseous element in the atmosphere, namely O, Si, Mg, Fe, Na, H, and K, which we refer to collectively as $X^g_i$ (where $i$ stands for any of the previous elements), as well as the total pressure ($P_{tot}$) of the atmosphere. 
For any combination of $(X^g_i, P_{tot}),$ the CEA/NASA code can provide the molecular composition of the atmosphere at chemical equilibrium in the gas. To couple the atmospheric model to the infinite magma ocean model, we must find the only atmospheric composition that simultaneously  satisfies (1) the liquid/vapor equilibrium, (2) the mass conservation of O, and (3) the mass conservation of H. The thermodynamic constraints are the following:
 \begin{itemize}
 \item Liquid/gas equilibrium states that the partial pressure of metal-bearing species evaporated from the melt (SiO, SiO$_2$, Na, K, Fe, Mg) depends on the atmospheric \textit{f}$\OT$. Therefore, the following relations must be obeyed for the gas for each reaction $j$ reported in Table 1.
 \begin{equation}
    P_{vapor_j}=\frac{K_j(T)a(liquid_j)}{P_{\OT}^{s_j}}
     \label{eq_partial_pressures_equil}
 ,\end{equation}
where $vapor_j$ stands for the vapor species that bears the metal in evaporation reaction $j$ (e.g., $Na_{(g)}$ in reaction $\#3$), $liquid_j$ is the corresponding liquid oxide ($NaO_{0.5}$ for reaction $\#3$), $s_j$ is the stochiometric coefficient (one-quarter in reaction $\#3$), $K_j(T)$ is the equilibrium constant of vaporisation reaction $j$, and $a(liquid_j)$ is the activity of liquid $\#j$, that is $a(liquid_j)=X_j\times\Gamma_j$, where $X_j$ is the mole fraction of $liquid_j$ in the magma ocean (assumed to be fixed) and $\Gamma_j$ is the activity coefficient of $j$ in the liquid due to nonideal mixing effects. Activity coefficients,  $\Gamma_j$, of all melt oxide components are interpolated using outputs of the VapoRock code \citep{Wolf_2022}.  
  
  When the atmospheric composition verifies all the relations for 1<j<6 given by Eq.~\ref{eq_partial_pressures_equil}, the gas is at thermodynamic equilibrium with the liquid.
 \item We assume that all metals and O in the atmosphere come from the vaporization of liquid oxides.
 This imposes a mass conservation relation between oxygen and all evaporated metals, which is governed by the stoichiometric relations of all reactions reported in Table 1. We simply get
 \begin{equation}
     X^g_\O=2X^g_{\Si}+X^g_{\Mg}+\frac{1}{2}X^g_{\Na}+\frac{1}{2}X^g_{\K}+X^g_{\Fe}
     \label{eq_mass_O2}
 ,\end{equation}
 where $X^g_i$ is the number of moles of atom $i$ in the gas.

\item  The initial total content of H (measured as a pressure) is fixed to a constant value, $P_\H^0$, meaning that Equation \ref{PH0_def} must also be verified.

\end{itemize}

\begin{table}[h!]
  \begin{center}
  \begin{threeparttable}
    
    \label{table:Table_reaction_liquid_gas}
    \begin{tabular}{c|c|c|c} \caption{Liquid-gas reactions}
     \textbf{$\#$} & \textbf{Reaction } & \textbf{s} & \textbf{ref  }\\
       \hline
            1 & $\SiO_{2(\ell)} \Leftrightarrow \SiO_{(g)}+1/2 \OT$ &  1/2 & Chase(1998)    \\
            2 & $\SiO_{2(\ell)} \Leftrightarrow \SiO_{2(g)}  $ & 0 & Chase(1998)     \\
            3 & $\Na\O_{1/2(\ell)} \Leftrightarrow \Na_{(g)}+1/4 \OT$ & 1/4   & Chase(1998) \\
            4 & $\K\O_{1/2(\ell)} \Leftrightarrow \K_{(g)}+1/4 \OT$ & 1/4  & Chase(1998)   \\
            5 & $\Mg\O_{(\ell)} \Leftrightarrow \Mg_{(g)}+1/2 \OT$ & 1/2  & Chase(1998)     \\
            6 & $\Fe\O_{(\ell)} \Leftrightarrow \Fe_{(g)}+1/2 \OT$ & 1/2  & Chase(1998)     \\
        \end{tabular}
        \begin{tablenotes}\footnotesize
        \item[] On the left-hand side are the liquid oxides and on the right the gas oxides. We note that for species Na$_2$O and K$_2$O, we have considered the oxide normalized by the number of metal atoms, as in  \cite{Sossi_2020} and \cite{sossi_etal2019}. The mole fraction of $\Na\O_{1/2}$ in melt is of $\sqrt{\Na_2\O}$. The same applies to K$_2$O. Values of all thermodynamic constants were taken from \cite{Chase-JANAF}. 
        \end{tablenotes}
  \end{threeparttable}
  \end{center}
  
\end{table}


We note that we assume that Fe is only present as FeO in the melt, meaning Fe$_2$O$_3$ is ignored, as is reasonable to expect in a magma ocean in equilibrium with H$_2$ gas. \\ 

The calculation is performed as follows:

\begin{itemize}
    \item Choose a magma composition ($X^l_i$), temperature, and a total hydrogen content ($P_\H^0$) .
    \item Determine the atomic composition of the atmosphere ($X^g_i$ for all atoms i) and the total pressure $P_{tot}$. For all sets ({$X^g_i$, $P_{tot}$}), the corresponding molecular composition of the atmosphere is computed using the CEA/NASA chemical equilibrium code. The only solution to this problem is the ($X^g_i$, $P_{tot}$) set for which the molecular composition of the atmosphere simultaneously solves Equations  \ref{PH0_def} to \ref{eq_mass_O2}.
    \item The search is done using  an iterative method, with a simplex minimization algorithm. The calculation goes on until Equations \ref{PH0_def} to \ref{eq_mass_O2} are solved up to a relative accuracy of better than $10^{-3}$ for each partial pressure involved.
    
\end{itemize}

For the atoms considered in this paper (Si, Mg, K, Na, Fe, O, H), the following gas species are included in the CEA/NASA thermolib table (and are the same as those found in JANAF table): Fe, FeO, H, $\mathrm{H_2}$,
$\mathrm{H_2O}$,
$\mathrm{H_2O_2}$,
$\mathrm{HO_2}$,
K, 
$\mathrm{K_2}$, 
$\mathrm{K_2O}$,  $\mathrm{K_2O_2}$,  $\mathrm{K_2O_2H_2}$, $\mathrm{KH}$,  $\mathrm{KNa}$,  $\mathrm{KO}$, $\mathrm{KOH}$,  $\mathrm{Mg}$, $\mathrm{Mg_2}$, $\mathrm{MgH}$,  $\mathrm{MgO}$,  $\mathrm{MgOH}$,  $\mathrm{Na}$,  $\mathrm{Na_2},  \mathrm{Na_2O}$,  $\mathrm{Na_2O_2}$,  $\mathrm{Na_2O_2H_2}$,  $\mathrm{NaH}$,  $\mathrm{NaO}$,  $\mathrm{NaOH}$,  $\mathrm{O}$,  $\mathrm{O_2}$,  $\mathrm{O_3}$,  $\mathrm{OH}$,  $\mathrm{Si}$,  $\mathrm{Si_2}$,  $\mathrm{Si_3}$,  $\mathrm{SiH}$,  $\mathrm{SiH_2}$,    $\mathrm{SiH_3}$,    $\mathrm{SiH_4}$,   $\mathrm{SiO}$,   $\mathrm{SiO_2}$.

\subsection{Validation against other thermochemical codes}

 Appendix \ref{Appendix_compo_pure_mineral} displays the composition we find for the vapor in the absence of H. This is a case that has been extensively investigated in the past (see e.g., \cite{Visscher_Fegley_2013, Ito_2015}) and more recently in the VapoRock \citep{Wolf_2022} and  LavAtmos codes \citep{LavAtmos_2022}. Our calculated partial pressures are in very good agreement with those calculated by existing codes (compare our Appendix to Figure 4 of \cite{Wolf_2022} and \cite{LavAtmos_2022}).

\subsection{Validation against molecular dynamics calculations}

However, our calculations are limited to species that are included in the thermodynamic tables, such as JANAF, which may not be complete.
For the sake of completeness, we compare our thermodynamic model against {\em ab initio} molecular dynamics simulations. For this, we employ the VASP package based on the planar augmented wavefunctions \citep{Kresse:1993} of the density functional theory. We consider pyrolite melt, with the bulk silicate Earth composition \citep{McDonough1995}: 1/2Na$_2$O.2CaO.3/2Al$_2$O$_3$.4FeO.30MgO.24SiO$_2$. This six-component composition was used previously to model the crystallization \citep{Caracas:2019} and the structure of the magma ocean \citep{Solomatova:2019}.

We performed simulations well inside the liquid--vapor dome at 3000~K and a density of 0.4 g/cm3, at which neither liquid nor vapor are stable as a single phase, but coexist as a mechanical mixture. The simulations are realized within the generalized gradient approximation in the Perdew-Wang-Ernzernhof formulation 
\citep{Perdew:1996}, spin-polarized, and corrected for the van der Waals interactions in the gas phase \citep{Grimme:2010}. 

We used five different initial configurations and ran simulations lasting between 50 and 136 picoseconds with a time step of 2 femtoseconds. For each simulation, we first build the entire interatomic connectivity matrix \citep{Caracas:2021ge}, which allows us to define the chemical species. At 3000~K, we obtain a vapor whose main components are: SiO, FeO$_2$, Na, O$_2$, Mg, FeO, O, and SiO$_2$. 
The major difference with our thermodynamic calculation (and all other published calculations; see Appendix \ref{Appendix_compo_pure_mineral}) comes from the presence of FeO$_2$ observed in the ab initio simulation and not considered in any thermodynamic calculation. Moreover, the FeO$_2$ component is present at all temperatures in the simulations. This may indicate that further work is needed to reconcile thermochemical models and ab initio molecular dynamics calculations.


%

\section{Results: Atmospheric composition for a magma ocean degassing in hydrogen}
\label{sec_atmo_compo}
\subsection{Composition above the magma ocean}
\label{sec_atmo_compo_above_MO}
Our calculations are performed between T~=~1800~K and 3500~K, and for total hydrogen pressures ranging from $10^{-6}$ to $10^{6}$~bar. At higher pressure, hydrogen could be metallic (> 1~Mbar), which is beyond the scope of our paper, as we focus on terrestrial to sub-Neptune planets, and ignore giant planets. In addition, at such high pressures, the ideal-gas approximation on which the present paper is based fails. Examination of the virial coefficients for common gases indicate the assumption of ideality breaks down above $\sim$10$^{3}$ to 10$^{4}$ bar. Refractory species such as Al and Ca may also be of interest but are not expected to comprise significant fractions of silicate atmospheres \citep[e.g.][]{Fegley_2016_steam_atmo}; neither have they been detected in exoplanetary atmospheres. As such, we do not consider them in this work other than to define the mole fractions of melt oxides in the BSE. 

\begin{table}[h!]
  \begin{center}
  \label{tab:Table_BSE_COMPO}
    \caption{Composition of the magma ocean corresponding to a BSE \citep{PALME20141}}
    
    \begin{tabular}{c|c|}
     \textbf{$\#$} \textbf{Oxide } & \textbf{Molar fraction} \\
       \hline
            $MgO$ & 0.468 \\
            $\mathrm{\SiO_2}$ & 0.388 \\
            $\mathrm{FeO}$ & 0.0579 \\
            $\mathrm{Na_1O_{1/2}}$ & 0.00578 \\
            $\mathrm{K_1O_{1/2}}$ & $3.42\times 10^{-4}$ \\
        \end{tabular}
  \end{center}
\end{table}

\begin{figure}\centering
   \includegraphics[width=.5\textwidth]{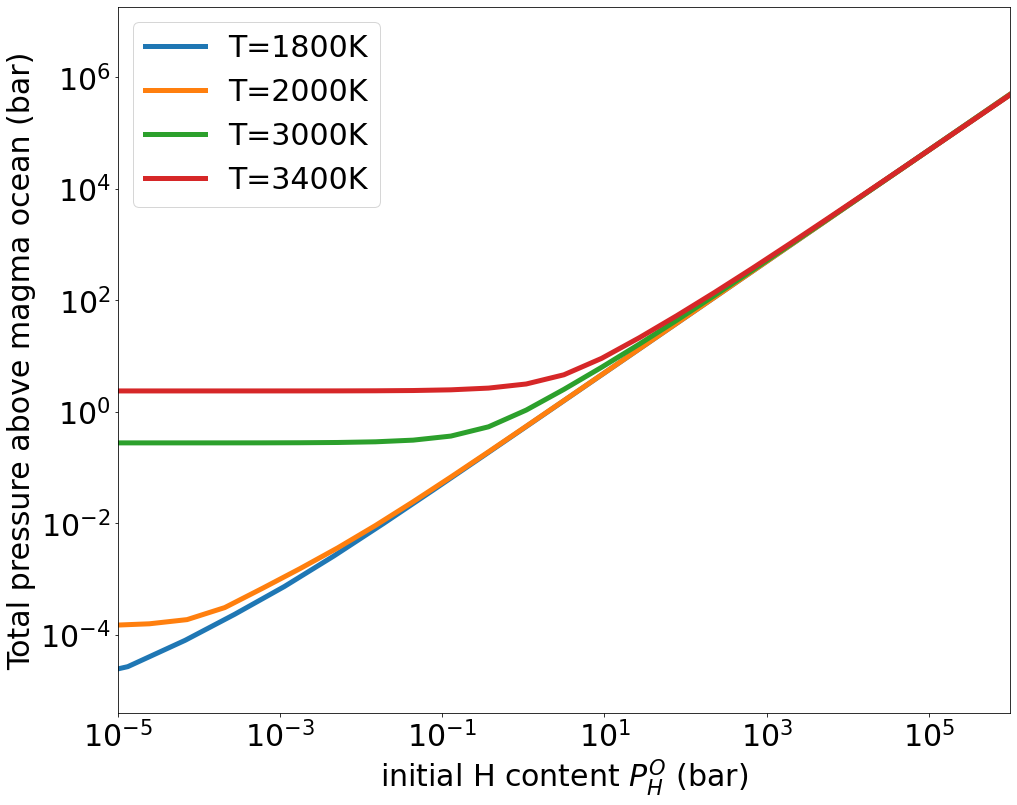}
   \caption{Total pressure of the gas at equilibrium with the magma ocean for different temperatures.}
    \label{Fig_total_pressure_above_MO}%
\end{figure}

\begin{figure}\centering
   \includegraphics[width=.5\textwidth]{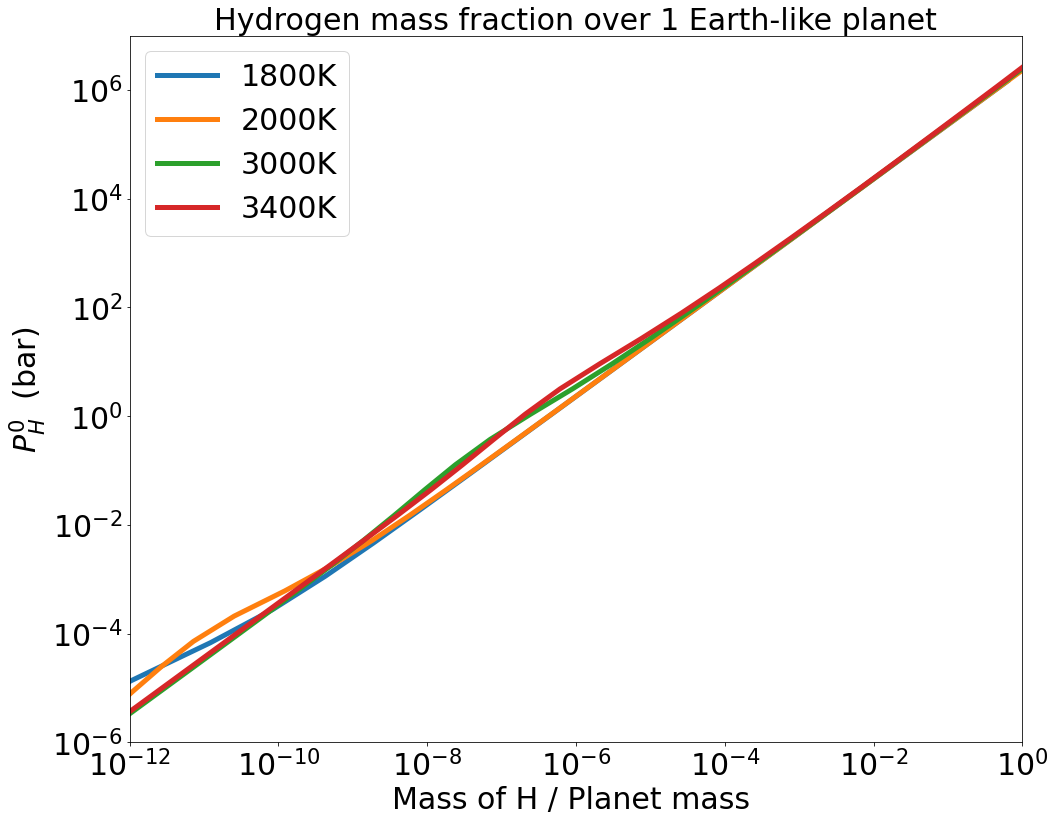}
   
   \caption{Total pressure of monoatomic hydrogen ($P_H^0$) versus planet mass fraction, assuming the planet has the mass and radius of the Earth.}
    \label{Fig_PH_Vs_mass}%
\end{figure}

\begin{figure}\centering
   \includegraphics[width=.5\textwidth]{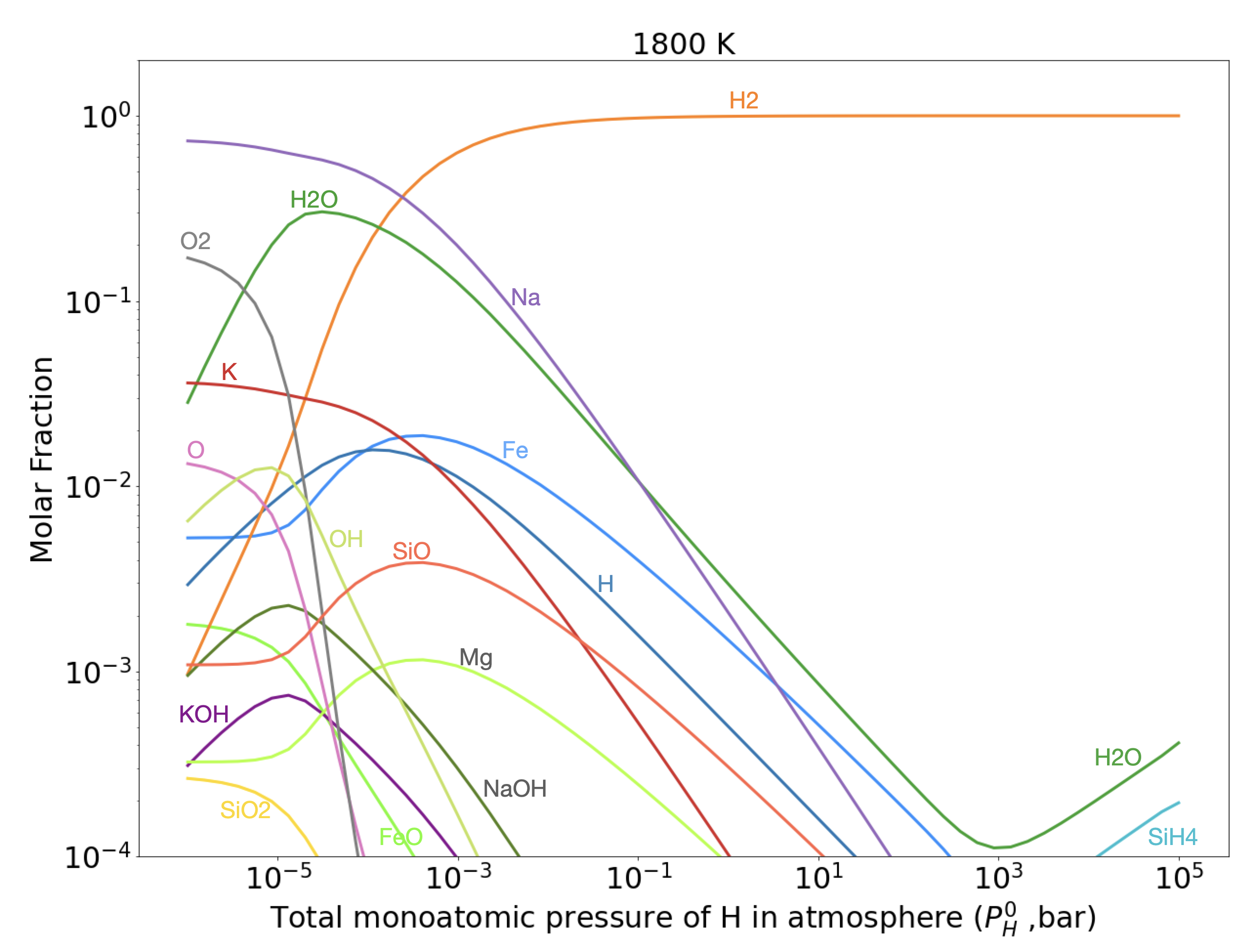}
   \caption{Molar abundances of most abundant species versus total monoatomic hydrogen pressure $P_\H^0$ for an atmosphere at equilibrium with a magma ocean at T=1800K. The same plot but with partial pressures is  provided in Appendix B. }
    \label{Fig_atmos_vs_PH_T=1800K}%
\end{figure}

\begin{figure}\centering
   \includegraphics[width=.5\textwidth]{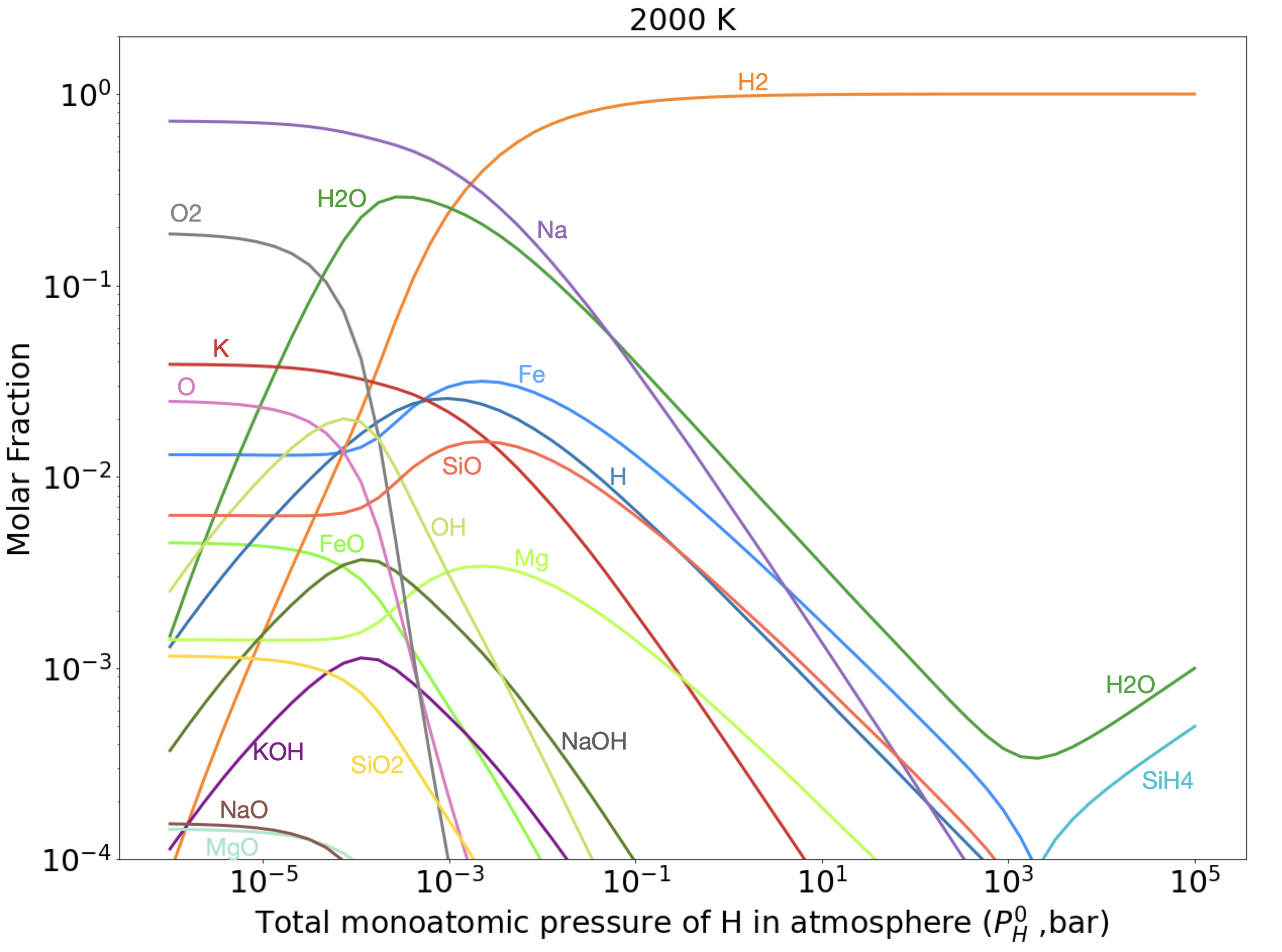}
   \caption{Molar abundances of most abundant species versus total monoatomic hydrogen pressure $P_\H^0$ for an atmosphere at equilibrium with a magma ocean at T=2000K. The same plot but with partial pressures is  provided in appendix B.}
    \label{Fig_atmos_vs_PH_T=2000K}%
\end{figure}

\begin{figure}\centering
   \includegraphics[width=.5\textwidth]{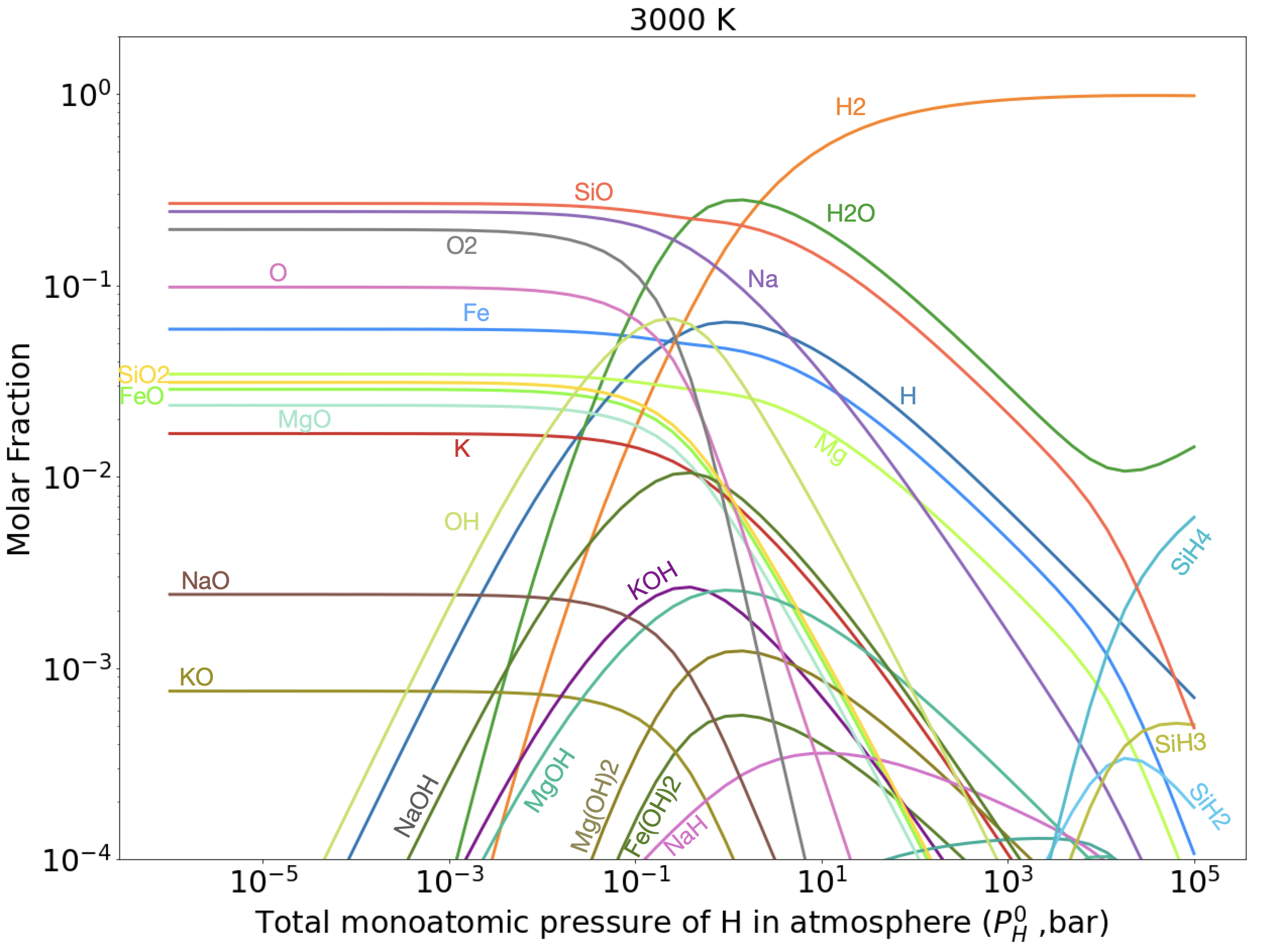}
   \caption{Molar abundances of most abundant species versus total monoatomic hydrogen pressure $P_\H^0$ for an atmosphere at equilibrium with a magma ocean at T=3000 K. The same plot but with partial pressures is  provided in appendix B.}
    \label{Fig_atmos_vs_PH_T=3000K}%
\end{figure}

\begin{figure}\centering
   \includegraphics[width=.5\textwidth]{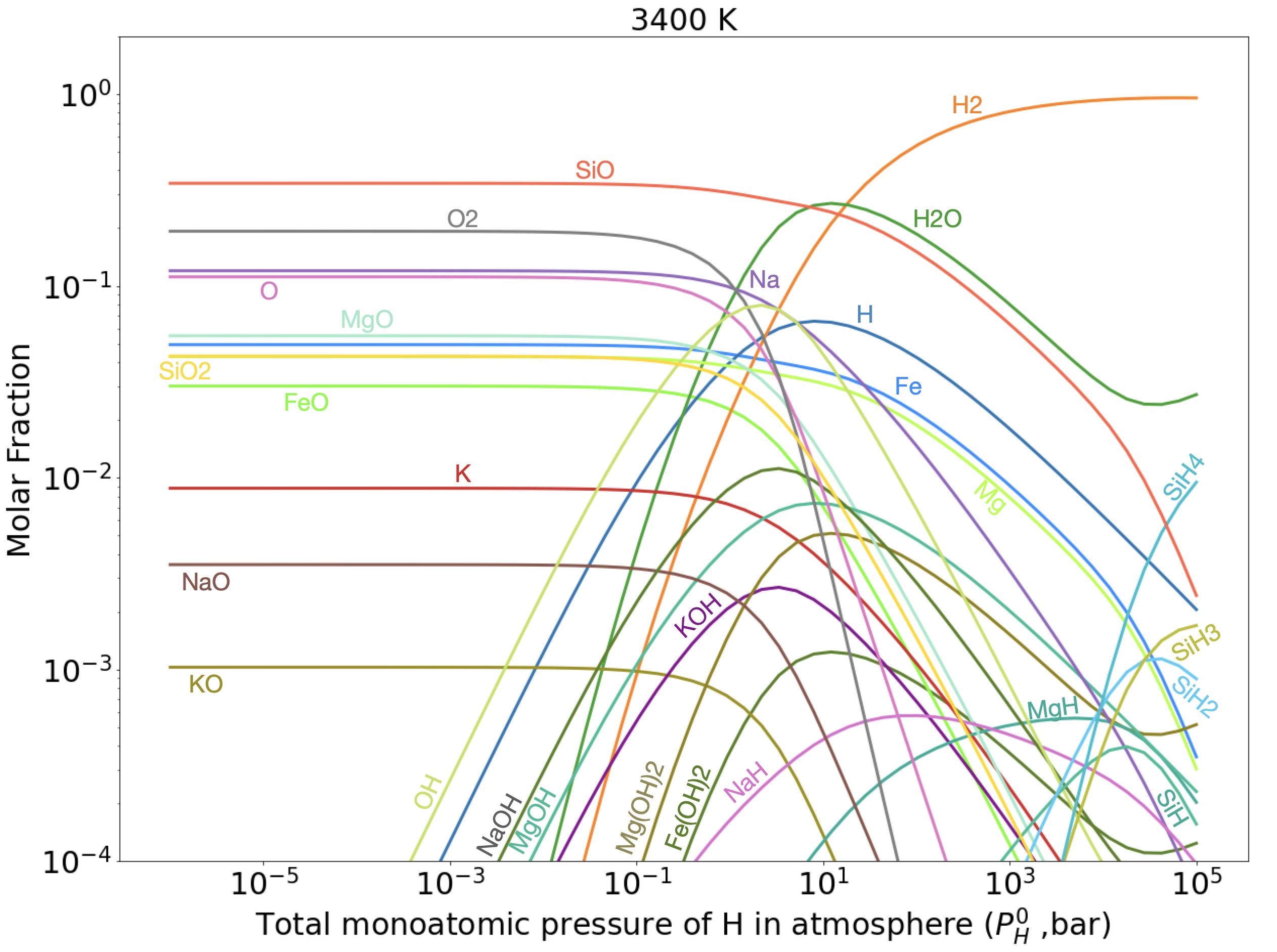}
   \caption{Molar abundances of most abundant species versus total monoatomic hydrogen pressure ($P_\H^0$) for an atmosphere at equilibrium with a magma ocean at T=3400K. The same plot but with partial pressures is  provided in appendix B.}
    \label{Fig_atmos_vs_PH_T=3400K}%
\end{figure}

Here, we present four fiducial calculations with magma oceans at 1800, 2000, 3000, and 3400~K and varying total H pressure.
The total pressure of the atmosphere above the magma ocean is plotted in Figure \ref{Fig_total_pressure_above_MO}. Figure \ref{Fig_PH_Vs_mass} provides the conversion between H pressure and mass of hydrogen (with respect to the planet's mass). The composition of the atmospheres at equilibrium with the magma oceans are plotted in Figs. \ref{Fig_atmos_vs_PH_T=1800K} to \ref{Fig_atmos_vs_PH_T=3400K}. A rapid comparison of these figures shows that the transition from a pure mineral atmosphere to a hydrogenated mineral atmosphere occurs when $P_H^O$ becomes comparable to the mineral vapor pressure (i.e., about $10^{-5}$, $10^{-4}$, $10^{-1}$, and 10 bars for T=1800, 2000, 3000, 3400~K, respectively). These pressures can be converted to a total mass of captured H using the following  relation (assuming constant surface acceleration at the planet's surface:$g=GM_p/R_p^2$):

\begin{equation}
    M_H=\frac{4\pi R_p^4}{GM_p}\sum_{i}\frac{P_i\nu^i_H\mu_H}{\mu_i}
,\end{equation}

where $M_p$, is the planet mass, $i$ is any molecular species in the atmosphere, $\nu^i_H$, $\mu_H$, and $ \mu_i$ are the stochiometric coefficient of H in molecule $i$, the molar mass of H, and the molar mass of molecule $i,$ respectively. Assuming an Earth-like planet, the conversion between $P_H^0$ and planet mass fraction is displayed in Figure \ref{Fig_PH_Vs_mass}.

We do see that for an Earth-like planet, amounts of H of $P_H^0=10^{-5}, 10^{-4}, 10^{-1}$, and $10$ bars correspond to planet mass fractions of $\sim 10^{-12}, 10^{-11},10^{-7}$, and $10^{-6} $, respectively. This demonstrates that a negligible amount of H can efficiently hydrogenate a mineral atmosphere. These mass fractions of H are several orders of magnitude smaller than the initial amount of captured H during the formation of the planet, which ranges from $10^{-3}$\% to 100$\%$ of the planet's mass \citep{Owen_2020_Hydrogren_planets}. 

Considering the molecular composition of the atmosphere  (Figs.
\ref{Fig_atmos_vs_PH_T=1800K}, \ref{Fig_atmos_vs_PH_T=2000K}, \ref{Fig_atmos_vs_PH_T=3000K}, \ref{Fig_atmos_vs_PH_T=3400K}), the same behavior is observed as in the simple two-reaction model  (section \ref{sec:toy_model}). As the H content increases (on the X axis), the partial pressures of gases derived from evaporation of the magma ocean  (those containing Si, Mg, Fe, and K atoms) increase by several orders of magnitude relative to the H-absent case (compare the left- and right-hand sides of Figs. \ref{Fig_atmos_vs_PH_T_PP=1800K} to \ref{Fig_atmos_vs_PH_T_PP=3400K}), a key result of the present paper. In Appendix B, we provide the atmospheric compositions represented with partial pressures.

In terms of molar fraction, for the T=1800~K case (Fig. \ref{Fig_atmos_vs_PH_T=1800K}), for low H contents ($P_\H^0 < 10^{-5} $bar) the atmosphere is dominated by Na,$\OT$ and K (in order of decreasing abundance), as in the classic rocky atmosphere case (see e.g., \cite{Schaefer_2012} ), and the total pressure is about $10^{-5}$ bar. However, for cases in which $P_\H^0 > 10^{-4}$~bar, the atmospheric composition becomes increasingly dominated by  $\mathrm{H_2}$ and $\mathrm{H_2O}$ as volatile species, and the dominant metal-bearing species are now Fe and Na (in order of decreasing abundance). Fe is efficiently released and increases linearly with $P_H^0$. For $P_H^0$=1 bar, NA is 100 times more abundant that in the H free case(see partial pressure plot in Appendix \ref{Fig_atmos_vs_PH_T_PP=1800K}), and Fr is 1000 times more abundant than in the H-free case . Interestingly, for $P_\H^0 > 10^3$~bar, a new gas species becomes dominant (just after $\HT$): $\SiHF$ ("Silane", a toxic gas with a strong repulsive odor) and $\SiHT$ ("Silanide"; 100-1000 times less abundant). 

At 2000~K, the behavior is qualitatively the same as for 1800~K.
While Na is the most abundant metallic species in the absence of H for T=1800 and 2000~K, it is always overtaken by Fe and SiO for H content larger than $10$~bar.

For T~=~3000~K (Fig.~\ref{Fig_atmos_vs_PH_T=3000K}), the same qualitative behavior is observed but more species are involved.
At this temperature, a pure rocky atmosphere (i.e., low H) is dominated by SiO, Na, O$_2$,O, Fe,SiO$_2$, FeO, Mg, and MgO, with a total pressure of about 0.1 bar. When $P_\H^0 \sim 0.1$~bar, the dominant volatile species become $\HT$ and $\HTO$, and the abundance of $\OT$ decreases. For $1 <P_\H^0 < 10^4$~bar, the dominant heavy species are SiO, Fe, Mg, and  Na. For $P_\H^0 > 10^4$~bar, SiO decreases sharply as silicon is converted into $\SiHF$ and $\SiHT$. In turn, the $P_{\OT}$ increases again for  $P_\H^0 > 10^4 $  bar because of the liberated oxygen in this process (see Fig. \ref{Fig_atmos_vs_PH_T_PP=3000K}).

Finally, for  T~=~3400~K (Fig.~\ref{Fig_atmos_vs_PH_T=3400K}), at low H content, the pure rocky atmosphere has a total pressure of about 3 bar, and is dominated by SiO, O$_2$,  O, $\SiO_{2(g)}$ , Na, Fe, and $\Mg\O_{(g)}$. 


For $P_\H^0 > 100$~bar, the atmosphere becomes dominated by $\HT$, $\HTO$, and SiO while the abundance of $\OT$ decreases.
For $P_\H^0 > 10^4$~bar, SiH$_4$ and SiH$_2$ become the most abundant heavy molecules. The shift of this transition to higher initial H contents is mostly due to the temperature dependence of reaction \ref{reac_H2O}, in which the pH$_2$/pH$_2$O ratio decreases with increasing temperature, such that the formation of the reduced species silane and silanide are delayed to higher total H contents in hotter atmospheres.


From the above considerations, the following conclusions can be drawn:
\begin{itemize}
    \item  Adding hydrogen to the evaporated atmosphere above the magma ocean increases the partial pressure of all evaporated species owing to the decrease in $f\OT$. This therefore promotes efficient evaporation of the magma ocean as more material goes into the atmosphere, and will also favor spectroscopic detection as the column density of evaporated species will be higher and the scale heights larger.

    \item The vapor becomes strongly hydrogenated (where $\HT$ and $\HTO$ become the dominant species and where the atmosphere chemistry deviates strongly for pure mineral atmosphere) when $P_\H^0 > P_{RA}$ where $P_{RA}$ is the pressure of the rocky atmosphere in the absence of H. Converted into a mass of atmospheric H, this represents from $10^{-12}$ to $10^{-6}$ times the planet mass only.
   
    \item The appearance of $\SiHF$ and $\SiHT$ occurs for $P_\H^0 > 10^3$~bar in every case (representing $10^{-3}$ times the planet mass of hydrogen for a Earth like planet) owing to association reactions; for example, SiO + 2H$_2$ = SiH$_4$ + H$_2$O, in which there are three moles of gas in the reactants compared with only two in the products. At these high H pressures, $\SiHF$ and $\SiHT$ become the most abundant heavy molecules (after $\HT$ and $\HTO$). 
\end{itemize}

\subsection{Signature of a hydrogenated magma ocean?}

It is interesting to speculate as to how we could detect the presence of a magma ocean beneath a hydrogen-rich atmosphere.  In the absence of hydrogen, the results presented above, and in previous studies \citep{Schaefer_2012,Schaefer_2010_atmo_formation,Lupu_2014_atmo_after_GI,Schaefer_2012,Ito_2015,Ito_Ikoma_2021, jaggi_etal2021, charnoz_etal2021}, show that Na should be released by a magma ocean for T~>~1500~K and should remain the dominant species up to 2500~K. At higher temperatures (>~2500~K), SiO should be the most abundant mineral species in the vapor. 

When H is present ---even in small quantities---, $\mathrm{H_2O}$, Fe, Na,  and SiO become dominant in the atmosphere. This suggests that the detection of $\mathrm{H_2O}$, Si, Na, or Fe for a rocky planet with an equilibrium temperature of $T_{eq}>2000K$ may point to the presence of a magma ocean below the atmosphere and the presence of a small amount of H.
As of today, heavy species, such as Si or Mg, have never been detected in the atmosphere of a rocky or sub-Neptune planet, whereas they have been detected in the atmospheres of giant exoplanets \citep{Zieba_2022}. In order to provide a framework for their detection in the future, below we report some atomic ratios that may be of interest for future observations designed to unveil the presence of a magma ocean.

\begin{figure}
   \includegraphics[width=.5\textwidth]{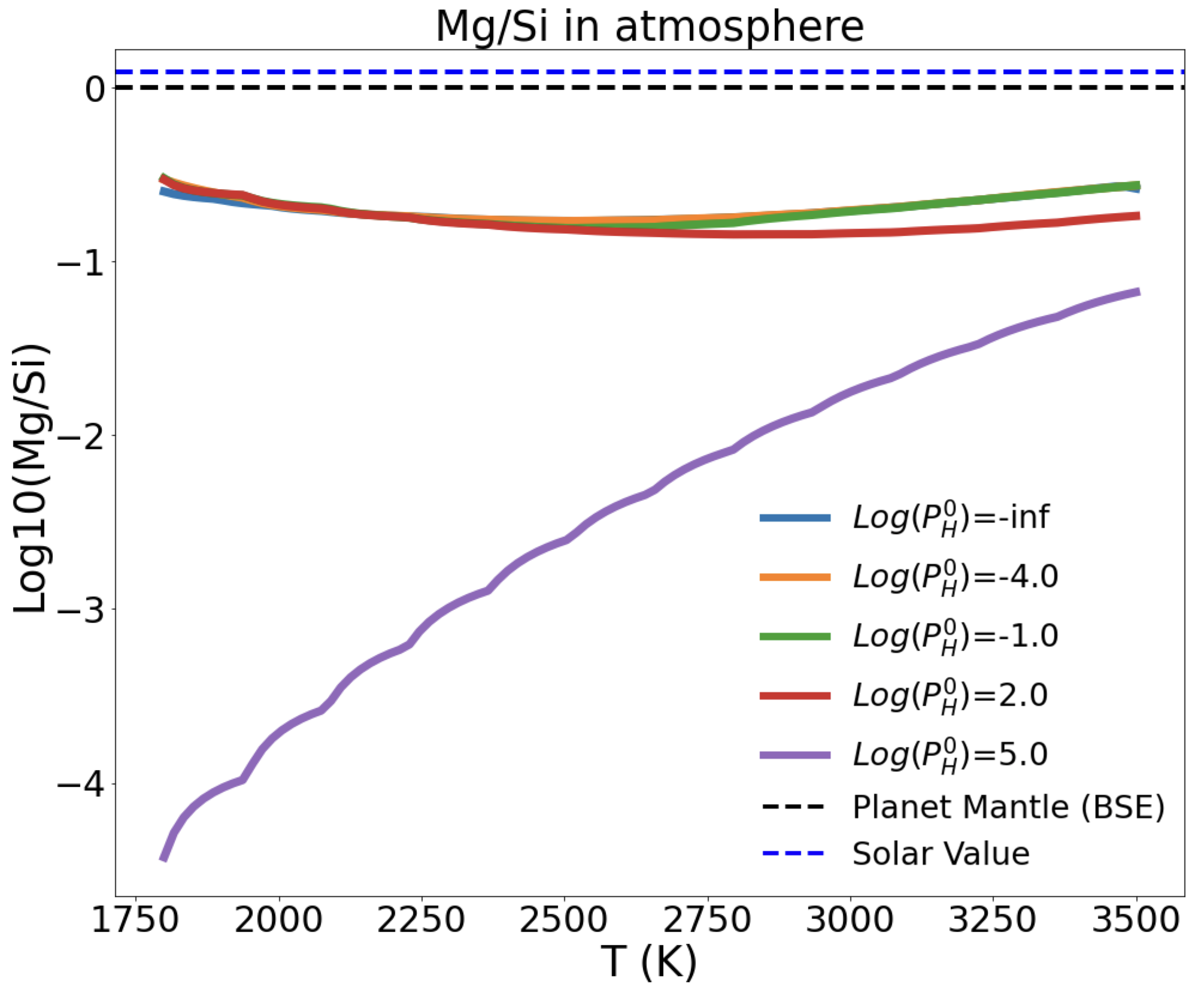}
   \caption{Molar Mg/Si ratio in the vapor above a magma ocean as a function of surface temperature. Colors display the amount of hydrogen in the planet (Log($P_\H^0$) in bars). The blue dashed line shows the solar value of Mg/Si and the black dashed line shows the BSE value.
   }
    \label{Fig_MG_SI}%
\end{figure}

The Mg/Si ratio in the atmosphere above the magma ocean is found to deviate strongly from that of the planet's mantle (Fig.\ref{Fig_MG_SI}). For $P_H^0$ $<$ 1000 bar, the atmospheric Mg/si ratio is almost constant at about $10\%$ of the mantle value. For higher hydrogen content, the Mg/Si ratio decreases  ---by several orders of magnitude---  below the values of both the star and the mantle. As Mg and Si are two major elements that condense at relatively high temperatures ($\sim$ 1350 K at 10$^{-4}$ bar) from the nebular gas, the fractionation of these elements by condensation or evaporation during planetary accretion should be limited, at least for stars with solar-like C/O ratios \citep{larimerbartholomay1979}. Therefore, assuming that the Mg/Si ratio of the planetary mantle is equal to that of the host star, their detection on an ultrahot rocky planet may provide insights into the magma ocean composition. A caveat associated with this conclusion is that Si remains lithophilic in the planetary mantle, that is, it does not enter the core in appreciable amounts, as is the case on Mercury for example \citep[e.g.,][]{malavergne_etal2010}.

\begin{figure}
   \includegraphics[width=.5\textwidth]{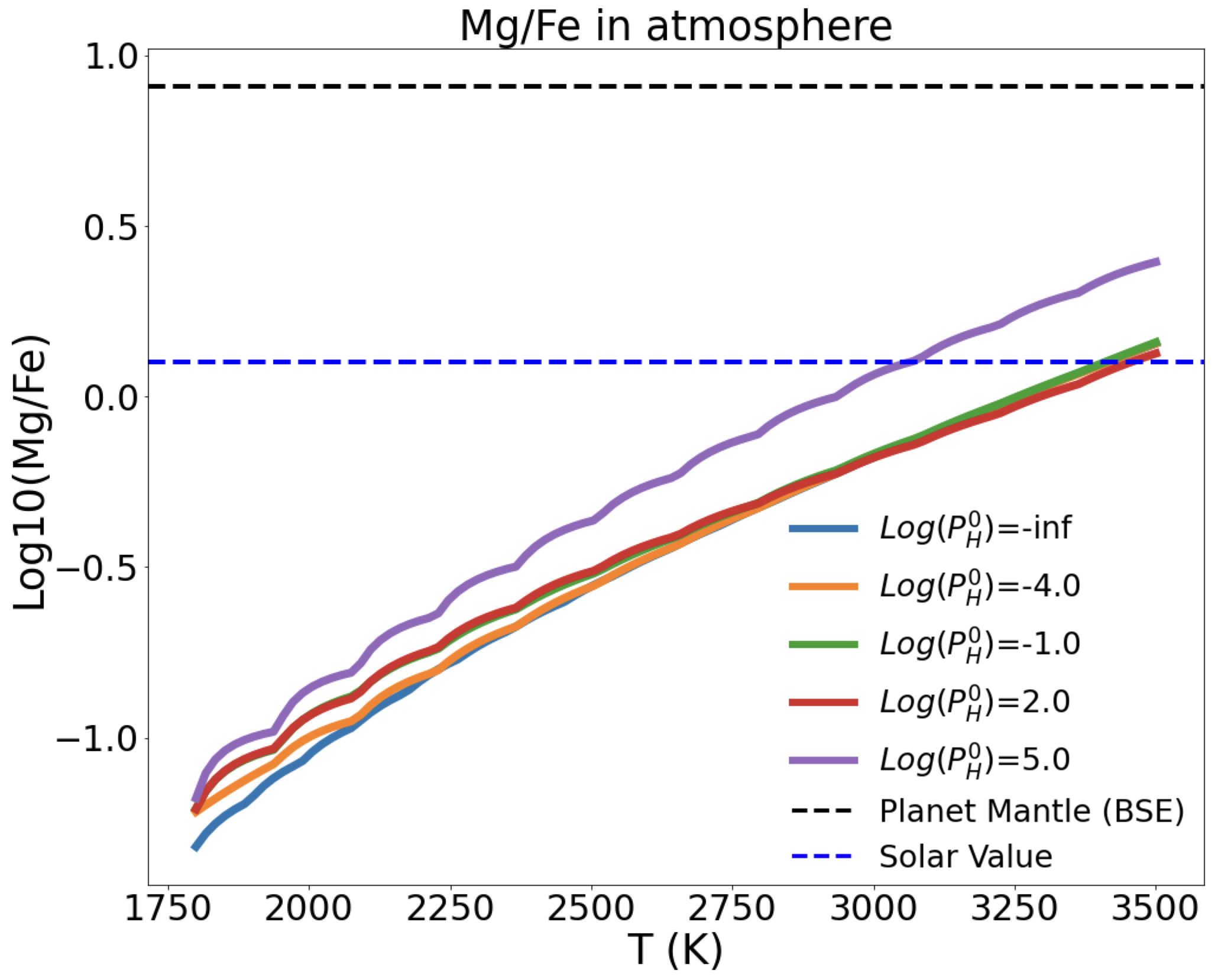}
   \caption{Molar Mg/Fe ratio in the vapor above a magma ocean as a function of surface temperature. Colors display the amount of hydrogen in the planet (Log($P_\H^0$) in bars). The blue dashed line shows the solar value of Mg/Fe and the black dashed line shows the BSE value.
   }
    \label{Fig_MG_FE}
\end{figure}

The $Mg/Fe$ ratio of the atmosphere is of interest because our models show that it is independent of H content (Fig. \ref{Fig_MG_FE}). This is because \textit{p}Mg and \textit{p}Fe are proportional to \textit{a}MgO and \textit{a}FeO in the liquid phase, and inversely proportional to \textit{f}O$_2^{1/2}$ (cf. Eq. \ref{eq_partial_pressures_equil}). Measuring the $Mg/Fe$ ratio in a molten exoplanet atmosphere may provide insights into the Mg/Fe mantle composition independently of the H content of the atmosphere.

\begin{figure}
   \includegraphics[width=.5\textwidth]{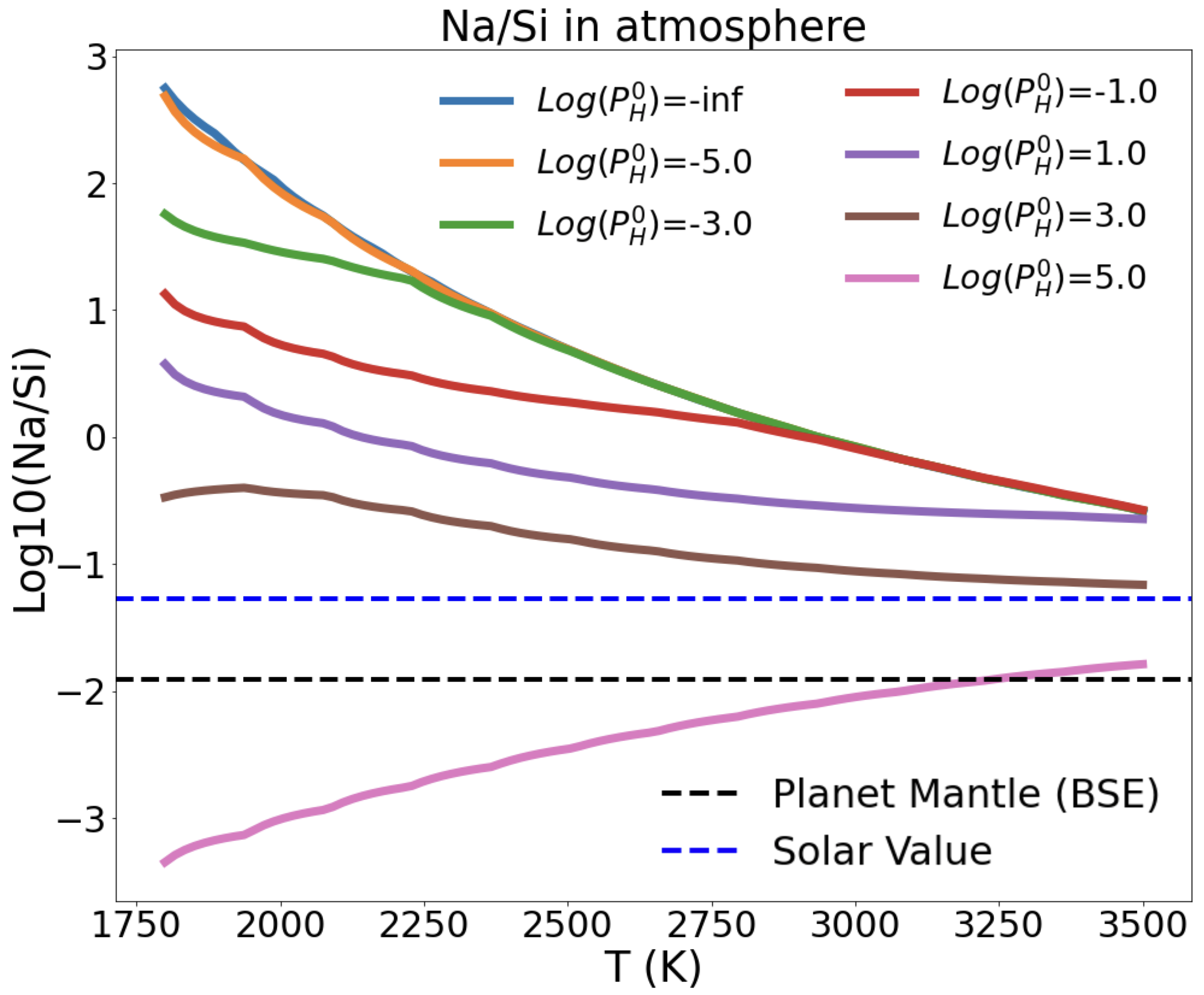}
   \caption{Molar Na/Si ratio in the vapor above a magma ocean as a function of surface temperature. Colors display the amount of hydrogen in the planet (Log($P_\H^0$) in bars). The blue dashed line shows the solar value of Na/Si and the black dashed line shows the BSE value.
   }
    \label{Fig_NA_SI}
\end{figure}

On the other hand, the Na/Si ratio of the atmosphere is strongly dependent on the H content of the planet (Fig. \ref{Fig_NA_SI}). If this ratio can be measured (or deduced based on SiO partial pressure, see \cite{Wolf_2022}), it may provided constraints as to the H content of the atmosphere. Such a conclusion is dependent upon Na still being present in the magma ocean today despite its tendency to evaporate and escape over the lifetime of the planet (see following section, and \cite{erkaev_etal2022}).

\section{Atmospheric escape }
Now that we are able to compute the atmosphere composition above the magma ocean, we wish to compute the efficiency of escape. This is a very complex problem, and has mostly been investigated for hydrogen-rich planets. The treatment of the full problem using an MHD-hydrocode including multiple species and complex chemistry (as we do here) is well beyond the scope of the present paper. Still, to obtain an approximate estimate of the escape efficiency of such an atmosphere, we developed a model inspired by the popular energy-limited model, which we present below.

\subsection{Method of calculation}
\label{section_atmo_escape_method}
We try to quantify the efficiency of atmospheric escape from planets with a long-lived magma ocean embedded in a primordial hydrogen envelope. We focus on planets very close to their star so that stellar heating is strong enough to maintain a magma ocean during the lifetime of the star. 
While it is well established that hot-Jupiter planets are massive enough to retain their atmospheres (see e.g., \cite{Valencia_2015}), Earth-sized planets should have lost their primordial H envelope in less than 10~Myr ( e.g., \cite{Valencia_2015,Lopez_2017,erkaev_etal2022}). 
What happens in the intermediate regime of super-Earths/sub-Neptunes is comparatively poorly understood. Of particular importance is the observation of a bimodal radius distribution of super-Earths, with a well-identified valley in the radius range 1.5-2~$R_{\oplus}$ (see e.g., \cite{Fulton_2017, Gupta_2019}).
It has been proposed that this "valley" marks the transition from the smaller rocky planets ("super-Earths") to planets containing a few per cent of the planet’s total mass in a H/He- rich envelope ("sub-Neptunes"; see e.g., \cite{Owen_WU_2013, Lopez_Fortney_2014, Ginzburg_et_al_2016,  Ginzburg_Sari_2017}). 
This transition may be due to rapid photoevaporation of highly irradiated H-rich atmosphere, which may not survive for planets with radii of $<1.5-2 R_{\oplus}$. In this section, we do not investigate the origin of this transition; rather, we study the consequence of the presence of a small amount of hydrogen on the efficiency of magma evaporation and escape of the atmosphere produced in such a scenario.


Here, we consider the case of an atmosphere in thermodynamic equilibrium with a magma ocean of constant temperature at its base as described in the previous section. Models of strongly irradiated planets show that escape is always efficient for terrestrial-mass planets \citep{Ito_2015,Benedikt_2020,Ito_Ikoma_2021} and that the particle velocities in the atmospheres of these planets far exceed those necessary for hydrostatic equilibrium. Although this conclusion is unlikely to hold for a cold atmosphere (which may indeed be hydrostatic), this assertion may be justified for ultrahot planets (T~>~1800~K) with irradiated atmospheres, depending on the average molar mass of the atmosphere. 
For heavy species (Si, Mg), \cite{Ito_Ikoma_2021} find strong atmospheric escape (with mass flux of about 0.4~$M_{\oplus
}/Gyr$) in coupled hydrodynamical and  radiative simulations, for an Earth-size planet at 0.02~AU devoid of hydrogen. These authors conclude that the atmosphere is never at hydrostatic equilibrium and is in "blow-off" regime, with a Parker-wind-type transonic velocity profile.
\cite{Konatham_2020_thermal_escape} argue that significant to catastrophic escape occurs if the atomic thermal velocities reach $>10\%$ of a planet's escape velocities. This  means that an Earth-like planet at a temperature of 3000~K  may lose all molecules up to 100 atomic mass units, whereas a planet of 
2 Earth masses may lose atoms up to gasses with molar weight up to $\sim 25$ atomic mass units. Similarly, \cite{Johnstone_2019_escape} find that even Ar (40~AMU) can escape from a strongly irradiated Earth-like planet (but these authors do not include heavier species). In the latter study, the H/N/O/C/He/Ar atomic ratios in the upper escaping atmosphere are found to vary between 0.75 and 2.5 relative to the same ratios in the lower atmosphere. As long as the atmosphere is in the hydrodynamic escape regime (blow-off), it seems that no strong atomic fractionation occurs during the escape, within a factor of a few \citep{Johnstone_2019_escape}. 

In all the studies mentioned above, the gravitational stratification of species is not found because diffusive timescales are always longer than advective timescales. This is why gravitational stratification is ignored here.



We note that cooling via vibrational bands of $\HTO$ \citep{Yoshida_2022} and atomic lines for N$_2$ and O$_2$ \citep{Nakayama_2022} has been found to be efficient enough to cool down the atmosphere and prevent escape from habitable planets under very strong XUV irradiation (up to about $1000$ times the present Earth value, which is relevant for very young stars). However, in the case we consider here, that is, ultrahot and ultrashort-period exoplanets, the XUV flux may be $ >10^4$ times  the value for Earth today owing to the very short stellar distance (0.01-0.02 au). In such extreme cases, \cite{Yoshida_2022} do find hydrodynamical escape.

A full treatment of the computation of atmospheric escape is beyond the scope of the present paper. Here, we limit our analysis to simple estimates based on the popular "energy limited" approximation, following the method of Salz et al. (2015), which provides a useful scaling law 
for energy-limited escape; this law is calibrated in 1D numerical simulations for irradiated, hydrogen-dominated planets. The mass flux (Kg/s) escaping from the atmosphere is approximated by:
\begin{equation}
    \dot M=\frac{-3 \beta^2 \eta F_{XUV} }{4 KG \rho_p},
    \label{equ_Mdot_escape}
\end{equation}

where $\beta=R_{XUV}/R_p$ is the ratio of the radius of absorption of the  radiation from an XUV star to the physical radius of the planet ($R_p$), and $\eta$ is a parameter standing for the "heating efficiency", which is the fraction of input radiative energy converted into thermal expansion of the atmosphere. $F_{XUV}$ is the incident stellar flux in the XUV band (in $J/s/m^2$) and K is a parameter accounting for the effect of stellar tides (which facilitate escape). Here, we use K=1 for simplicity. 

Numerous physical mechanisms are subsumed into $\eta$ and its value is highly dependent on the stellar spectrum, the gravity of the planet, and the atmospheric composition (see e.g., \cite{Valencia_2015, Salz_2016, Ito_Ikoma_2021}). \cite{Salz_2016} find that $\beta$<2 and  depends sensitively on the planet's gravitational potential well and the  flux of the star. The following parametrization is proposed \citep{Salz_2016}:

\begin{equation}
    \begin{split}
    Log_{10}(\beta)=Max[0., -0.185Log_{10}\left(\frac{\Phi_G}{erg/g} \right)+ \\
    0.021Log_{10}\left(\frac{F_{XUV}}{ erg/cm^2/s}\right)+2.42],
    \label{Eq_Saltz_parameter_BETA}
    \end {split}
\end{equation}

where $\Phi_G=GM_p/R_p$. Numerical simulations by \cite{Salz_2016}, which are valid for a H$_2$-dominated atmosphere, show that for planets with a deep gravitational well (i.e., large $\Phi$), escape is difficult and $\eta$ may be $\ll~1$ (giant planets), whereas for planets with shallow gravitational potential (i.e., small $\Phi$), escape is relatively efficient  (Earth- and super-Earth-mass planets). Following \cite{Salz_2016} and noting $\nu =Log_{10}\left(\frac{\Phi_G}{erg/g} \right)$ $\eta_H$
for a hydrogen-dominated atmosphere: 
\begin{equation}
 Log_{10}(\eta_H)=
    \begin{cases}
      -0.5-0.44(\nu-12.) & \text{ for $12.<\nu<13.11$ }\\
      -0.98-7.29(\nu-13.11) & \text{for $13.11<\nu<13.6.$ }
    \end{cases}
\end{equation}

The above fit, which is valid for H$_2$-dominated atmospheres, may not be valid for pure mineral atmospheres.  \cite{Ito_Ikoma_2021} find that, in pure mineral atmospheres, atmospheric cooling is much more efficient than in H$_2$-dominated atmospheres, resulting in low escape efficiency with  $2\times 10^{-4}<\eta<4 \times 10^{-3} $, which is much lower than $0.1 < \eta < 0.5 $ for H-dominated atmospheres.  To take this compositional effect into account, we replace $\eta$ in Eq.~\ref{equ_Mdot_escape} by:

\begin{equation}
    \eta_{eff}=\eta_H\times \left(\frac{1.- e^{-x}}{1.-e^{-1}}\right)+5\times10^{-4},
    \label{eq_heta_effectif}
\end{equation}

where $x$ is the atomic fraction of H in the atmosphere. Although not optimal, the above prescription ensures that H$_2$-dominated atmospheres escape at the efficiency given by \cite{Salz_2016}, whereas mineral atmospheres escape with an efficiency in the (lower) range of the values reported by \cite{Ito_Ikoma_2021}.

Finally, concerning the time-dependent XUV flux at the location of the planet, we adopt the fit to XUV flux data reported for young stars by \cite{Valencia_2015}: 

\begin{equation}
    F_{XUV} (J/m^2/s)=
    \begin{cases}
       29.7\times 10^{-3}(0.1)^{-1.23}\left(\frac{1}{a^2}\right) \text{ for t < 0.1~Gyr} \\
       
       29.7\times 10^{-3}\left(\frac{t}{1~gyrs} \right)^{-1.23}\left(\frac{1}{a^2}\right)   
       
       \text{ for t > 0.1~Gyr.}
    \end{cases}
    \label{equ_FXUV_time}
\end{equation}

The calculation is performed as follows. We start with a planet mass $M_p$ and radius $R_p$ with surface acceleration $g$, surface temperature $T,$ and mass of H in the atmosphere $M_H$, corresponding to an initial pressure of monoatomic H: $P_\H^0=M_H*g/(4\pi {R_p}^2)$.
For a given $T$ and $P_\H^0$ , the atmospheric molecular composition above the magma ocean is computed as described in section \ref{sec_method_full_model}. We compute the total mass loss according to Eq.~\ref{equ_Mdot_escape}. We assume the star is 5~Gyr old. Following the two preceding arguments regarding inefficient gravitational stratification in the blow-off regime, the total mass flux of each atom is $\dot M\times X_i$, where $X_i$ is the atomic mass fraction of atom $i$ in the escaping gas.

We do not consider here the dissolution of $\mathrm{H_2}$ or $\mathrm{H_2O}$ in the magma ocean as this will be the subject of a future study. We note that the dissolution of $\mathrm{H_2}$ and $\mathrm{H_2O}$ into the magma ocean may lead to the constitution of a reservoir of H and O, which may lengthen the duration of H escape \citep{hiermajumder2017, bower2022retention}. Consequently, the timescales of H escape reported in the following section should be considered as lower bounds. In addition, the substantial solubility of H as $\mathrm{H_2O}$ in the magma ocean \citep{sossi2023solubility} indicates that it would be largely ($\geq$ 95 \% of its total budget) dissolved should the entire mantle of an Earth-mass planet remain molten. However, as the magma ocean cools, the solubility of $\mathrm{H_2O}$ exceeds the mass fraction of available liquid in which to dissolve, leading to late-stage release of significant amounts of O and H (as H$_2$O) into the atmosphere when the atmospheric pressure is sufficiently low \citep{Solomatova:2021a}. This may serve to increase the \textit{f}O$_2$ relative to that initially imposed by the H$_2$-rich nebular gas, and thus modify atmospheric chemistry. We warn the reader that these effects are not considered here. 

\subsection{Results}

\begin{figure}\centering
   \includegraphics[scale=0.3]{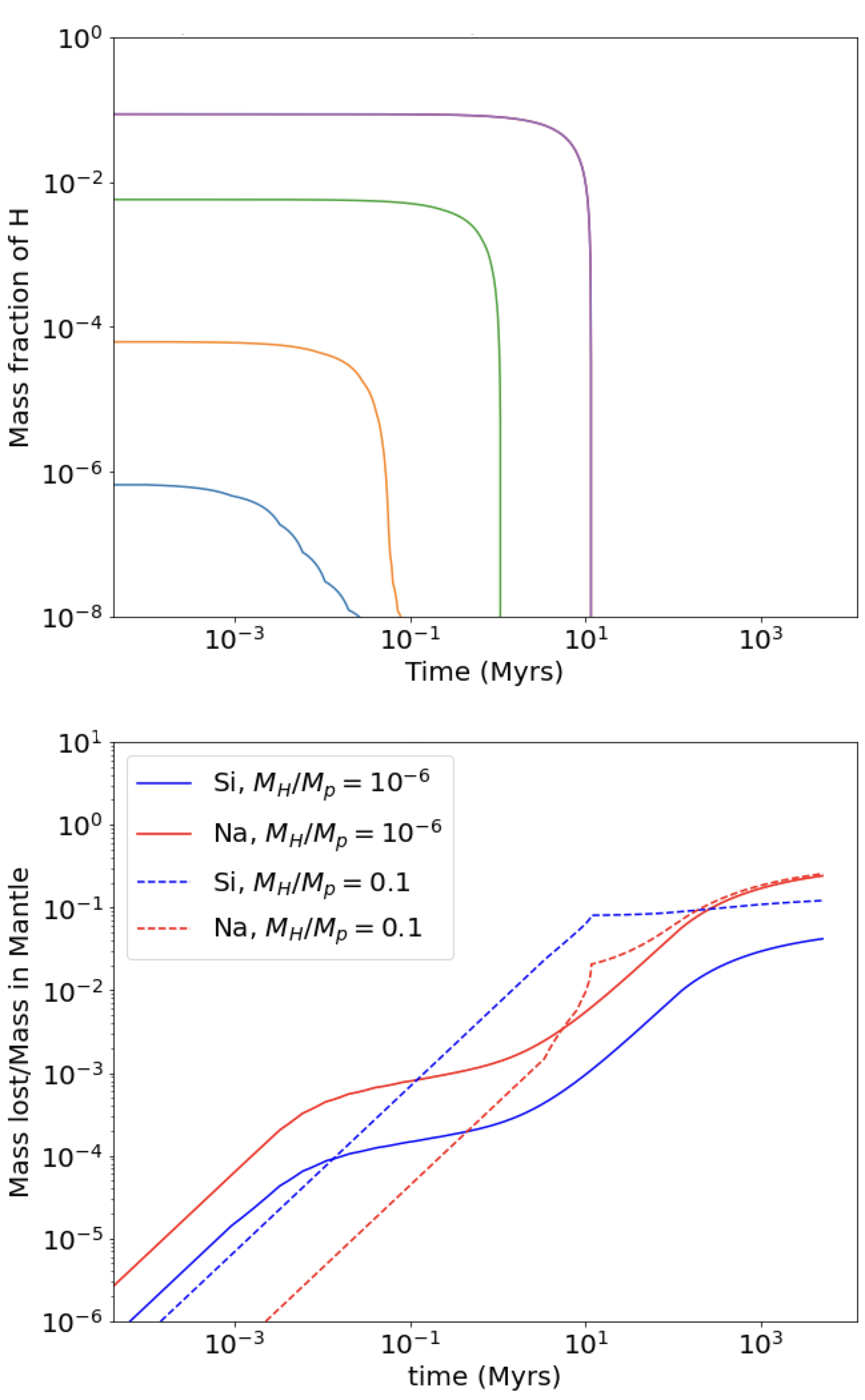}
   \caption{Escape from a 1 $M_{\oplus}$ and $R_{\oplus}$ planet located at 0.02 au from its star. Top: Mass of H vs time for different initial mass fractions of hydrogen. Bottom: Mass fraction of lost Si(red) and Na(blue) vs. time. The solid line shows these relations for an initial H mass fraction of $10^{-6}$ times the planet mass, and the dashed line is for an initial H mass fraction of 0.1 times the planet mass. }
    \label{Fig_HSiNA_vs_time_1Me}%
\end{figure}

Figure \ref{Fig_HSiNA_vs_time_1Me} displays the mass of hydrogen as a function of time for an Earth-like planet ($M_p=1~M_{\oplus}$, $R_p=1~R_{\oplus}$) located at 0.02~au, starting with different hydrogen content from $10^{-6}$ to $10^{-1}$ times the planet mass. The top panel shows that most H is lost within about 100~Kyr to 10~Myr, depending on $M_H^0/M_p$. The bottom panel displays the lost mass of Na  (a proxy for the moderately volatile elements) divided by its mass in the planet's mantle magma ocean (in red; we assume to first order that the magma ocean is $50\%$ of the planet mass).
For $M_H^0/M_p=10^{-6}$ (red solid line), Na is lost at a roughly constant rate as long as hydrogen is present and the star is~<~10~Myr old when its XUV flux is intense. When hydrogen has disappeared, the efficiency of Na escape increases. This result may be surprising, but can be reconciled with the fact that, for high H contents, the Na mole fraction is very low; that is, the atmosphere is dominated by SiO or Fe (see previous section). In contrast, when the abundance of H drops to zero, the mole fraction of the atmosphere comprised of H increases by three orders of magnitude, meaning the escape rate of Na increases by a proportionate factor. Over gigayear timescales, the loss of Na is dominated by loss during the last 1 billion years and we find, at the end, that about $30\%$ of the mantle Na content is lost.

The role of H in influencing the Si content of the planet's mantle is more marked. The blue solid line in Fig. 14 shows Si evolution for $M_H^0/M_p=10^{-6}$. After 5~Gyr, the planet devoid of H has lost only $\sim$ $1\%$ of its Si. Conversely, for planets starting with hydrogen contents of $10 \%$ of the mass of the entire planet, about $10 \%$ of the total budget of Si is lost after 5~Gyr. This strong increase in Si loss with H content reflects the fact that when H is abundant, SiO becomes the dominant metal-bearing gas species.  In addition, the presence of H increases the efficiency of atmospheric heating ($\eta~>~0.1$, see section~\ref{section_atmo_escape_method}) and decreases the mean molar mass of the atmosphere.
We highlight the fact that, due to the simplicity of the models, the above values may not be understood as exact solutions, but rather as order-of-magnitude estimates of the extent of escape. Indeed, the values could be somewhat underestimated owing to the conservative approach that was taken in estimating the quantities used in section~\ref{section_atmo_escape_method}, though it should be mentioned that the composition of the magma ocean is assumed to remain fixed rather than evolve, as would be expected for a finite mass undergoing Rayleigh (fractional) vaporisation. 

\begin{figure*}
   \centering
   \includegraphics[scale=0.25]{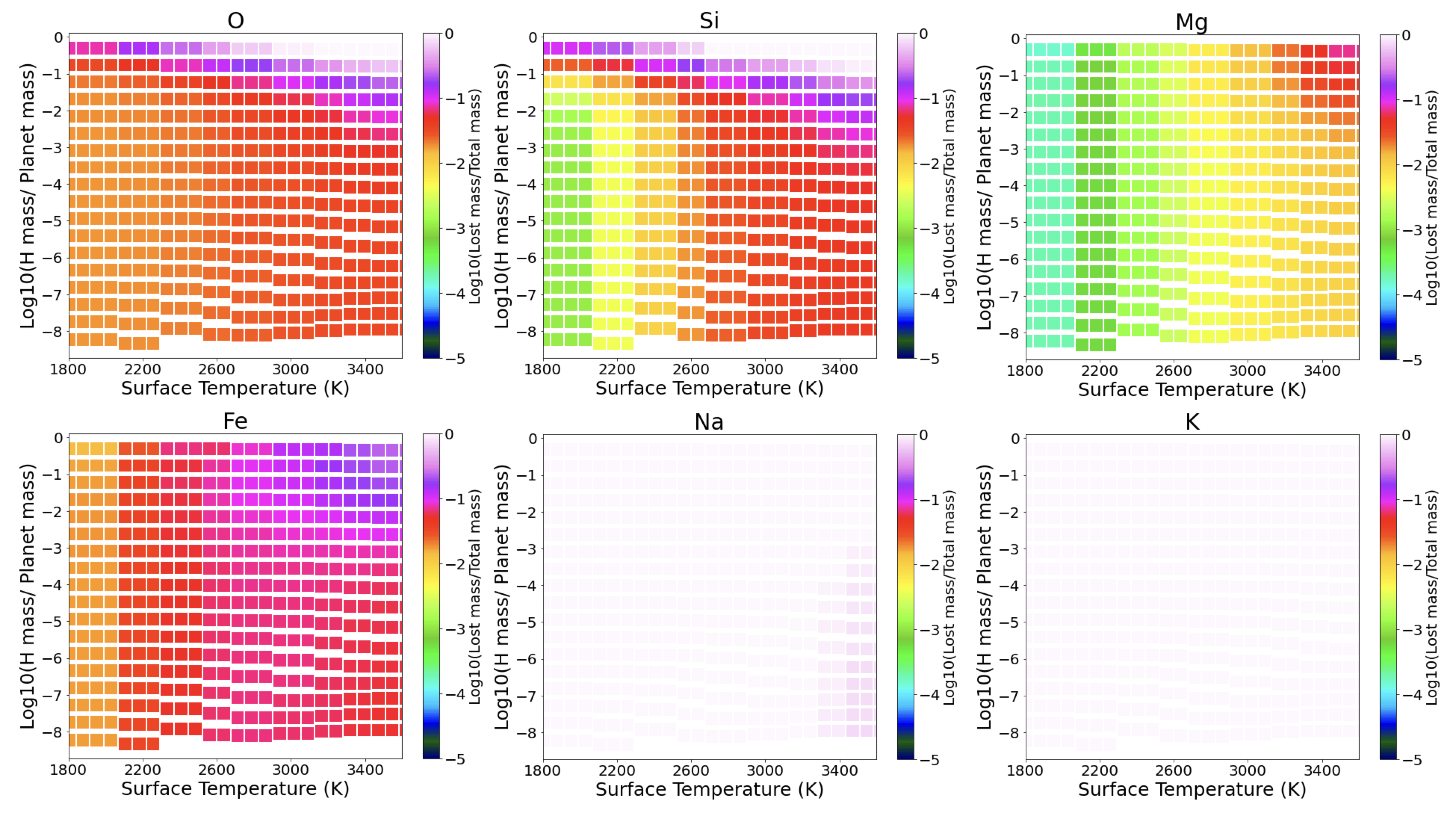}
   \caption{Each figure shows a color plot  of the mass fraction of an atom that is lost for temperature T (X axis) and initial hydrogen mass (Y axis) of the planet. The planet's solid core here is 1 $M_{\oplus}$ and 1 $R_{\oplus}$.}
    \label{Fig_Map_compo_escape_1M}%
\end{figure*}

We also explored different planet masses and radii and find that the main factors limiting the amount of metals and metal-bearing oxides and hydrides able to escape is the lifetime of the H envelope (controlled by the planet's density) and the surface temperature.  
For a more complete view of the effect of escape on each atmospheric constituent, Fig.~\ref{Fig_Map_compo_escape_1M} shows the lost mass fraction of every atom (O, Si, Mg, Fe, Na, K) for all combinations of temperature and initial mass fraction of hydrogen and for an Earth-like planet located at 0.02~au. The color of the plot indicates the mass fraction of each atom that is lost (with respect to the content in the mantle). White denotes that all atoms in the mantle are lost, purple indicates a loss of about 10$\%$, and so on.  The most volatile elements, Na and K, are completely lost for $T_{surf} < 3500$~K, which is in line with the results of \cite{Ito_Ikoma_2021}. 
For $T_{surf}> 3500$~K, the efficiency of Na and K loss is decreased because their molar fraction in the atmosphere is supplanted by other species (Si, Fe, Mg), thus diminishing their net loss in comparison to other species. Indeed the mass flux (eq. \ref{equ_Mdot_escape}) only depends on the F(XUV) flux and the planet mass and radius. In addition, the heating efficiency is low (<~0.001), such that when temperature is~>~3200~K and H content is low, the species that escape are mostly those that predominate in the atmosphere, that is, mostly Si, O, and Mg. 
If we now turn to Si,
we see that more than 10$\%$ of the planet's mantle Si is lost for $T_{surf}>2600$~K and for a H mass fraction of $M_H/M_p > 0.1 $. The same result applies for oxygen. As oxygen is the most abundant species by mass (about $50\%$ of the mass of Earth's mantle), this implies significant removal of the planet's mantle and an increase in average density (an effect that is not taken into account here). Concerning Fe, it is substantially lost (>10$\%$)  for temperatures > 2600K  and $M_H^0/M_p > 0.01 $ due to its high abundance in the vapor combined with its low abundance (6.3 mol \%) in the mantle.

Magnesium shows an  intermediate behavior between those of Si and Na, with most efficient loss occurring for $T_{surf}>3200$~K  and $M_H^0/M_p> 10^{-2}$. Up to $8 \%$ of Mg can be lost from the mantle in such extreme conditions. The less efficient loss of Mg stems from its lower volatility than Si and Fe for the BSE composition considered here. 

Therefore, in conclusion, we do see that for extreme surface temperatures (> 2600~K) and initial H mass content of > 1$\%,$ a significant fraction of the mantle's content ($\sim$ a few 10$\%$) could be evaporated during the planet's evolution. We note that  the results of this section must be taken with care because of the simplicity of the model and the numerous simplifications we had to introduce. The energy-limited escape is known to overestimate the escape, and the escape efficiency is not well known for atmosphere heavier than solar composition. In addition, vibrational and atomic radiative cooling lines during the escape \citep[see e.g.,][]{Yoshida_2022, Nakayama_2022} may act against escape. We therefore conclude that our simple model suggests that escape of heavy molecules may be made easier by the presence of H, but further investigation of this process is required  to confirm this finding.

\label{section_atmo_escape_results}

\section{Conclusions}
     We have studied the composition and the escape of an atmosphere above a magma ocean planet in the presence of hydrogen. In the present study, a special effort was dedicated to computing the atmospheric composition at thermodynamical equilibrium with the magma ocean, taking into account the effect of H on the atmospheric chemistry, and its retro-action on the liquid--atmosphere equilibrium. Our main results are summarized below:
     
    \begin{itemize}
    \item Hydrogen present in the atmosphere has a dramatic effect on the vapor composition. For T=2000~K and 3000~K, the atmosphere composition is strongly modified for atmospheric hydrogen contents of about $10^{-8}$ and  $10^{-6}$ times the planet mass, respectively (assuming an Earth mass planet) and becomes dominated by $\mathrm{H_2}$ and $\mathrm{H_2O}$. 
    
    \item Hydrogen lowers the O$_2$ partial pressure (=oxygen fugacity $fO_2$)  to make H$_2$O,  which happens in proportion to the mass of H added. As a consequence, the addition of hydrogen promotes efficient degassing of metals Na, K, Fe, Mg, and Si. These elements all evaporate according to reactions, $M^{x+s}O_{(x+s)/2}(l) = M^{x}O_{x/2}(g) + s/4O_2(g)$ in which the stoichiometric coefficient, s, is positive. Therefore, a small amount of hydrogen (comparable to the total pressure of the evaporated elements, and nearly negligible compared to the planet's mass) greatly enhances the extraction of these elements from the magma to the vapor phase.
    
    \item The atomic abundances  of heavy (i.e., rock-forming) species in the atmosphere may diverge significantly from the mantle value, and are controlled by the magma--atmosphere equilibrium.
    The Mg/Si ratio in the atmosphere is constant and independent of H content for T>3000~K and may be directly related to the composition of the planet's mantle. Conversely, the Na/Si  ratio is very sensitive to H content (owing to the different dependence of the partial pressures of their major gas species on $fO_2$, and may be useful in constraining the presence of H). Finally, the Mg/Fe ratio is independent of H and may be linked to the mantle composition. This suggests that spectroscopic observations of magma ocean planets could provide a means to constrain the mantle composition, even in the presence of hydrogen.
    
    \item For Earth-sized planets of < 0.02 AU and with a temperature above the magma ocean, that is > 2000~K, most H is lost in < 10 Myr, regardless of its speciation.
    
    \item For Earth-sized planets of < 0.02 AU and with a temperature above the magma ocean, that is > 3000~K, we find  significant loss of Fe, Si, and O (> 10$\%$ of their budget in the planet's mantle). In such extreme conditions, this may lead to a decrease in the  mass of the mantle, and in turn to an increase in the mean density of the planet (due  to core/mantle ratio decreases). This may potentially explain the high densities of strongly irradiated rocky planets.

    \end{itemize}

In conclusion, we show that the presence of relatively small fractions of H in the atmosphere may enhance the evaporation of metals from ultrahot rocky planets covered by a magma ocean, and, at very extreme temperatures, may lead to significant loss of heavy atoms such as Si, O, or Fe. Whereas no heavy species have yet been detected in rocky planetary atmospheres, we identify ratios of rock-forming elements in the vapor phase that are potentially able to reveal the presence and nature of a magma ocean, and subsequently probe the  composition of the planet mantle. In contrast to some recent studies \citep{Zilinskas_2022}, we do find that the presence of H strongly increases the partial pressure of major degassed species (see Figures \ref{Fig_atmos_vs_PH_T_PP=1800K} to \ref{Fig_atmos_vs_PH_T_PP=3400K}). This will significantly impact the resulting atmospheric spectra and (P,T) structure of the exoplanet. This will be the subject of a future study.

In the future, we will study the effect of dissolution of $\HT$ and $\HTO$ on the atmospheric composition of magma ocean exoplanets. We anticipate that the effect of dissolution of $\HTO$  will be to oxidize the atmosphere, thus lowering the reducing effect of H, which may lead to less efficient release of metallic species (boosted by H$_2$); it will also create an additional reservoir of H, probably resulting in an extended escape time of H. This will be studied in a future paper.

On a more technical note, we also emphasize that improvement of ab initio calculations and comparison with tabulated thermochemical data will be useful in the future, as ab initio calculations reveal the formation of a great diversity of species (sometimes not available in common data tables such as JANAF or CEA-NASA) and at higher temperatures than are typically accessible experimentally.

    \begin{acknowledgements}
      SC, AF, PT acknowledge financial support by LabEx UnivEarthS (ANR-10-LABX-0023 and ANR-18-IDEX-0001) and by the CNES (Centre National d'Études Spatiales). PT would also like to acknowledge and thank the ERC for funding this work under the Horizon 2020 program project ATMO (ID: 757858). RC acknowledges support from the European Research Council under EU Horizon 2020 research and innovation program (grant agreement 681818 – IMPACT to RC), the Research Council of Norway with project number 223272 and project HIDDEN 325567, and access to supercomputing facilities via eDARI stl2816 grants, PRACE RA4947 grant, Uninet2 NN9697K grant. PS thanks the Swiss National Science Foundation (SNSF) via an Ambizione Fellowship (180025), an Eccellenza Professorship (203668) and the Swiss State Secretariat for Education, Research and Innovation (SERI) under contract number MB22.00033, a SERI-funded ERC Starting Grant '2ATMO'. We thank the anonymous reviewer for their insightful comments that improved the quality of the paper.
\end{acknowledgements}

\bibliographystyle{aa} 
\bibliography{biblio} 

%
%

%
\appendix
\section{Composition of vapor in the absence of H}
\label{Appendix_compo_pure_mineral}
\begin{figure}[!h]\centering
   \includegraphics[width=.5\textwidth]{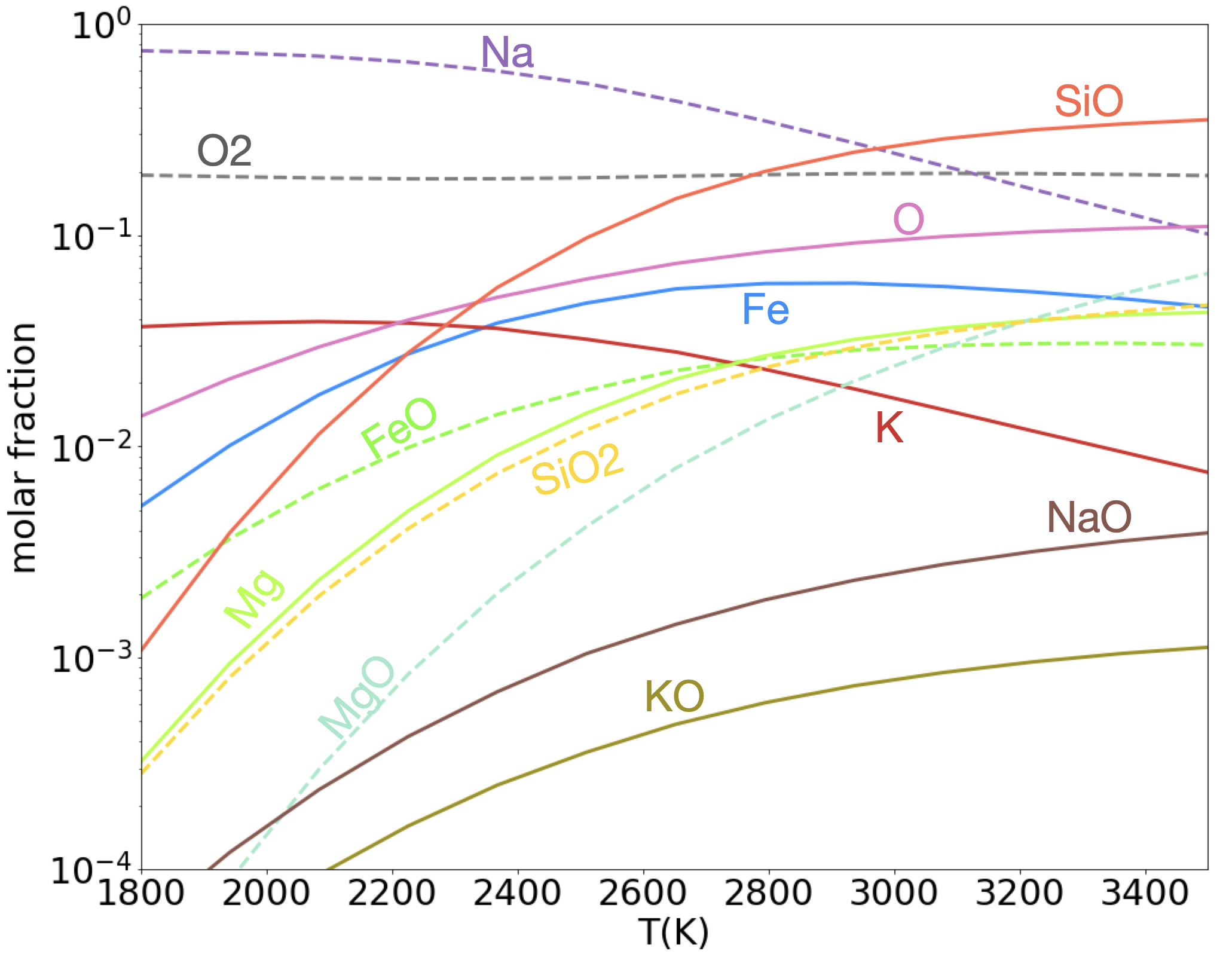}
   \caption{Molar fractions in the vapor at equilibrium with a magma with BSE composition, and without hydrogen.}
    \label{Fig_appendix_vapor_compo_noH}%
\end{figure}

Figure~\ref{Fig_appendix_vapor_compo_noH} shows the vapor composition at equilibrium with a liquid with BSE composition, computed with the procedure described in \citep{charnoz_etal2021}; it compares very well with composition reported in \cite{Ito_2015} \cite{Visscher_Fegley_2013} \cite{LavAtmos_2022}.

\section{Partial pressures in the hydrogenated atmosphere}

\begin{figure}[!h]\centering
   \includegraphics[width=.5\textwidth]{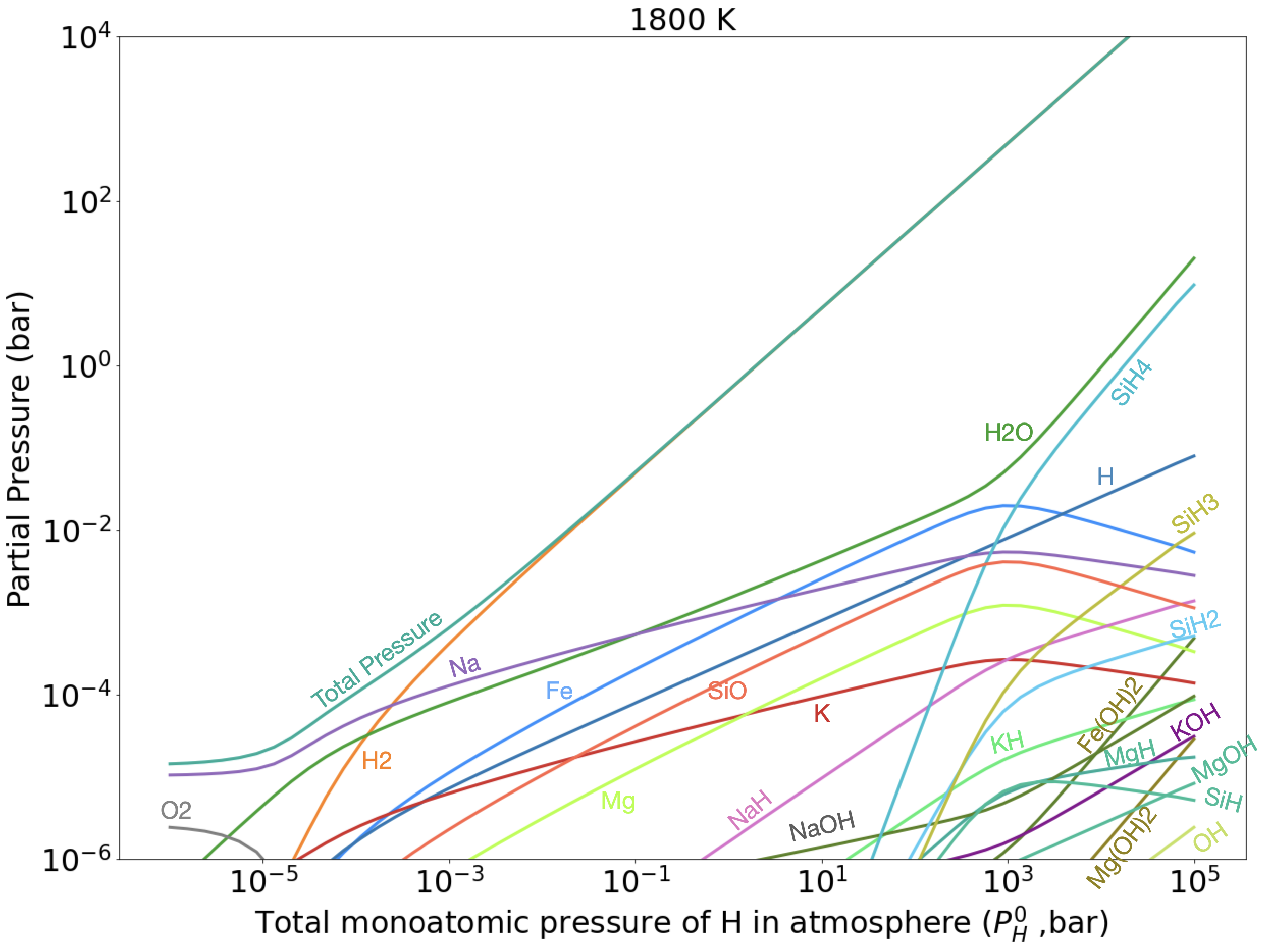}
   \caption{Molar abundances of most abundant species versus total monoatomic hydrogen pressure $P_\H^0$ for an atmosphere at equilibrium with a magma ocean at T=1800 K.}
    \label{Fig_atmos_vs_PH_T_PP=1800K}%
\end{figure}

\begin{figure}[!h]\centering
   \includegraphics[width=.5\textwidth]{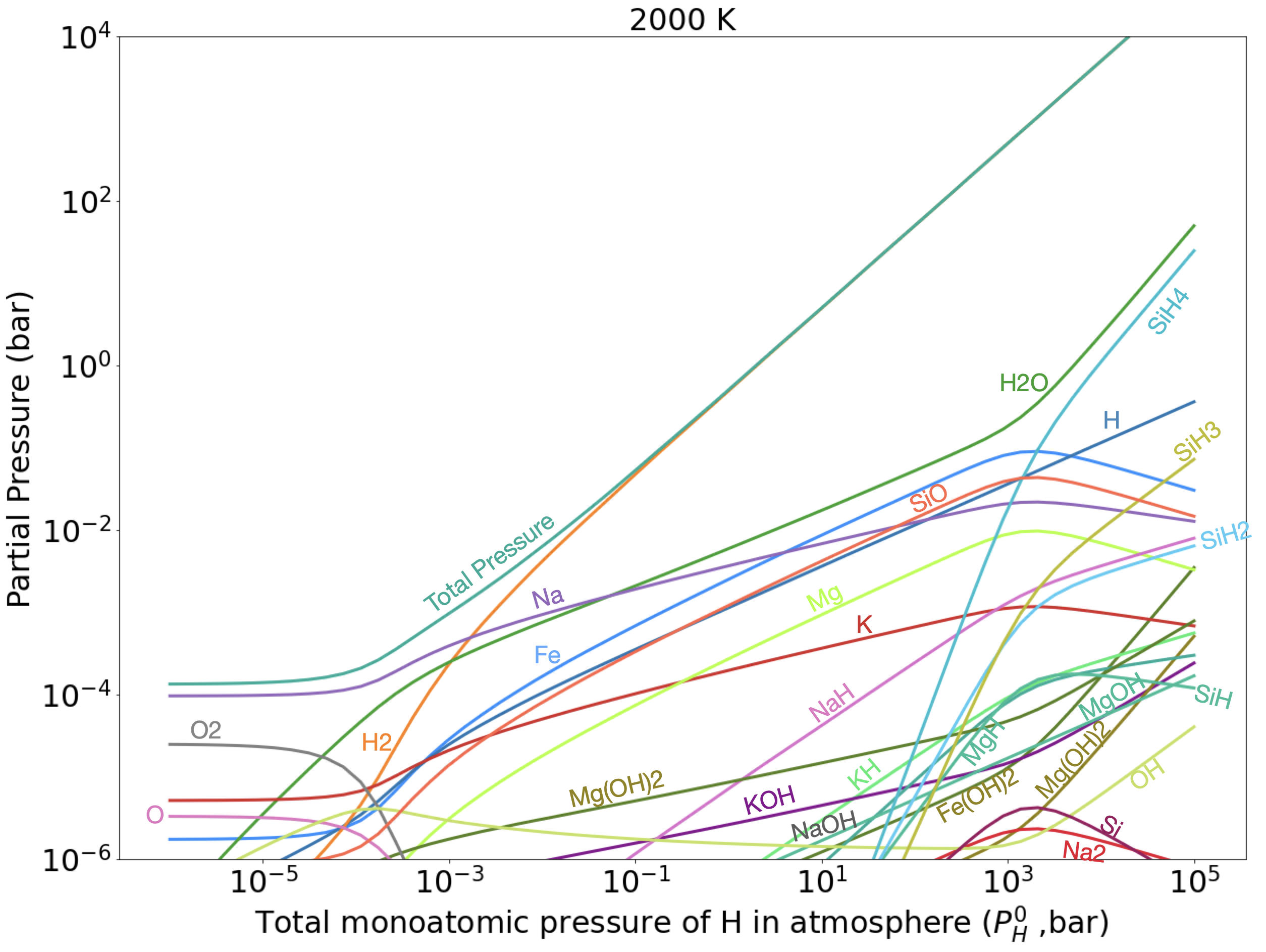}
   \caption{Molar abundances of most abundant species versus total monoatomic hydrogen pressure $P_\H^0$ for an atmosphere at equilibrium with a magma ocean at T=2000 K.}
    \label{Fig_atmos_vs_PH_T_PP=2000K}%
\end{figure}

\begin{figure}[!h]\centering
   \includegraphics[width=.5\textwidth]{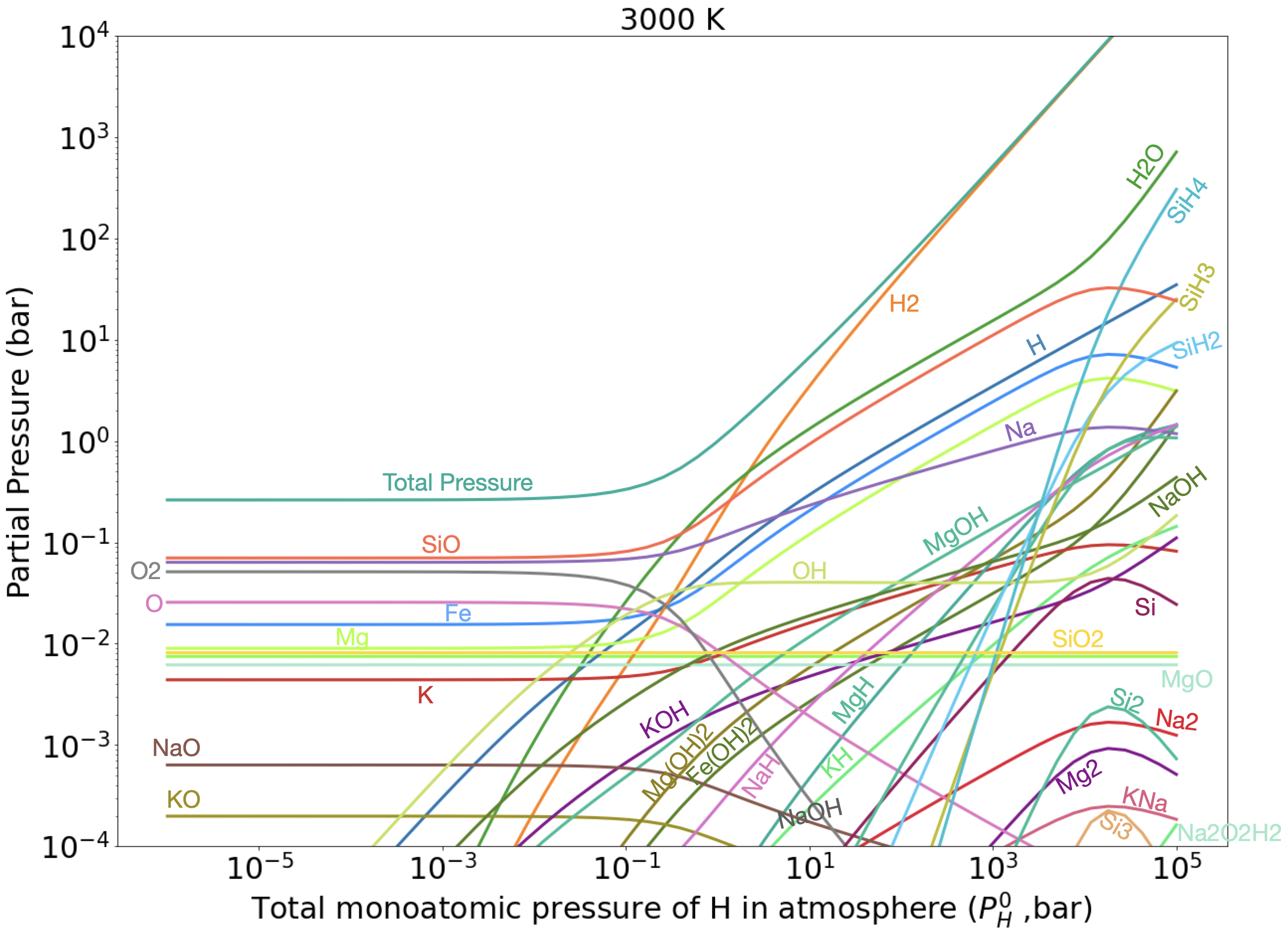}
   \caption{Molar abundances of most abundant species versus total monoatomic hydrogen pressure $P_\H^0$ for an atmosphere at equilibrium with a magma ocean at T=3000 K.}
    \label{Fig_atmos_vs_PH_T_PP=3000K}%
\end{figure}

\begin{figure}[!h]\centering
   \includegraphics[width=.5\textwidth]{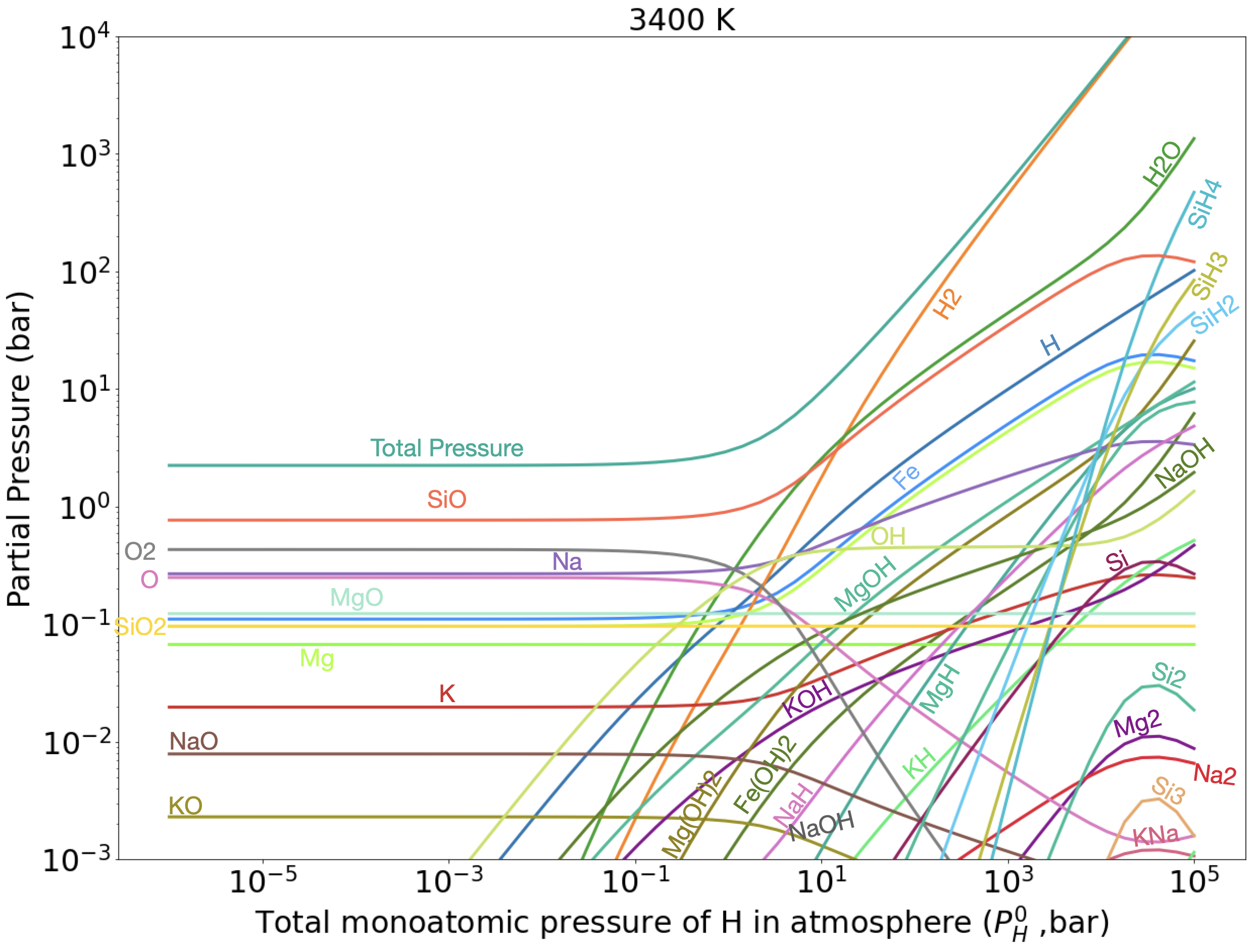}
   \caption{Molar abundances of most abundant species versus total monoatomic hydrogen pressure $P_\H^0$ for an atmosphere at equilibrium with a magma ocean at T=3400 K.}
    \label{Fig_atmos_vs_PH_T_PP=3400K}%
\end{figure}
\end{document}